\documentclass[fleqn,usenatbib]{mnras}

\usepackage{newtxtext,newtxmath}

\usepackage[T1]{fontenc}
\usepackage{ae,aecompl}

\usepackage{graphicx}	
\usepackage{amsmath}	
\usepackage{amssymb}	

\title[Chemical evolution of the MW disc]{Chemical evolution of the Milky Way: constraints on the formation of the thick and thin discs}

\author[Palla et al.]{
M. Palla$^{1,2}$\thanks{E-mail: marco.ball94@gmail.com},
F. Matteucci$^{1,3,4}$,
E. Spitoni$^{5}$,
F. Vincenzo$^{6,7}$,
V. Grisoni$^{8,3}$
\\
$^{1}$Dipartimento di Fisica, Sezione di Astronomia, Universit{\'a} degli Studi di Trieste, via G. B. Tiepolo 11, I-34131, Trieste, Italy\\
$^{2}$IFPU - Institute for Fundamental Physics of the Universe, Via Beirut 2, I-34014, Trieste, Italy\\
$^{3}$INAF, Osservatorio Astronomico di Trieste, via G. B. Tiepolo 11, I-34131, Trieste, Italy\\
$^{4}$INFN, Sezione di Trieste, via A. Valerio 2, I-34100, Trieste, Italy\\
$^{5}$Stellar Astrophysics Centre, Department of Physics and Astronomy, Aarhus University, Ny Munkegade 120, DK-8000, Aarhus C., Denmark\\
$^{6}$Center for Cosmology and AstroParticle Physics, The Ohio State University, 191 West Woodruff Avenue, Columbus, OH 43210, USA\\
$^{7}$Department of Astronomy, The Ohio State University, 140 West 18th Avenue, Columbus, OH 43210, USA\\
$^{8}$SISSA - International School for Advanced Studies, via Bonomea 265, I-34136, Trieste, Italy
}

\date{Accepted 2020 August 09. Received 2020 August 04; in original form 2020 June 18}

\pubyear{2020}

\begin{document}
\label{firstpage}
\pagerange{\pageref{firstpage}--\pageref{lastpage}}
\maketitle

\begin{abstract}
We study the evolution of Milky Way thick and thin discs in the light of the most recent observational data. In particular, we analyze abundance gradients of O, N, Fe and Mg along the thin disc as well as the [Mg/Fe] vs. [Fe/H] relations and the metallicity distribution functions at different Galactocentric distances. We run several models starting from the two-infall paradigm, assuming that the thick and thin discs formed by means of two different infall episodes, and we explore several physical parameters, such as radial gas flows, variable efficiency of star formation, different times for the maximum infall onto the disc, different distributions of the total surface mass density of the thick disc and enriched gas infall. Our best model suggests that radial gas flows and variable efficiency of star formation should be acting together with the inside-out mechanism for the thin disc formation. The timescale for maximum infall onto the thin disc, which determines the gap between the formation of the two discs, should be $t_{max}\simeq 3.25$ Gyr. The thick disc should have an exponential, small scale length density profile and gas infall on the inner thin disc should be enriched. We compute also the evolution of Gaia-Enceladus system and study the  effects of possible interactions with the thick and thin discs. We conclude that the gas lost by Enceladus or even part of it  could have been responsible for the formation of the thick disc but not the thin disc.
\end{abstract}



\section{Introduction}
\label{s:intro}

One of the fundamental constraint to study the formation and chemical evolution of our Galaxy are the abundance gradients along the Galactic thin disc. Abundance gradients have been observed in most of local spiral galaxies and show that the abundances of metals decrease outward from the galactic centre (\citealt{Belfiore19}).

A good agreement between observational properties and models for the Milky Way (MW) is generally obtained by assuming that the Galaxy forms by infall of gas (e.g. \citealt{Chiosi80,Matteucci89,Chiappini97,Spitoni11,Snaith15,Prantzos18}). Generally, the formation timescale of the thin disc is  assumed to be a function of the Galactocentric radius: this leads to an inside-out scenario (i.e. larger timescales for larger radii) for the disc build-up (e.g. \citealt{Matteucci89,Chiappini01,Schonrich17}) which helps in reproducing the observed gradients.\\
However, even if gas infall is important, the inside-out mechanism alone is not enough to explain the gradients observed in the Galactic disc (e.g. \citealt{Spitoni11,Mott13,Grisoni18}). \citet{Chiappini01} and \citet{Colavitti09} suggested to add a threshold in the gas density for the star formation (SF) or a variable star formation efficiency (SFE), i.e. higher in the inner regions than in the outer ones. Also radial gas flows had been suggested to fit the disc constraints at the present time (e.g. \citealt{Portinari00,Spitoni11,Bilitewski12,Cavichia14}). This explanation has a physical link with the MW formation via gas infall: in fact, the infalling gas may have a lower angular momentum than that of the gaseous disc, and the mixing between these two components may induce a net radial inflow (\citealt{Spitoni11}). Also other physical phenomena can induce radial gas flows (either inflow or outflows: e.g.  \citealt{Lacey85,Bertin96}). Anyway, these phenomena have generally small flow speeds and their effects on the gradients are assumed to be negligible (\citealt{Portinari00}).\\

However, present-day gradients are just part of the challenge to understand the formation and the evolution of the MW galaxy.\\
Recently, many surveys have been developed to study the MW, such as for example Gaia-ESO (\citealt{Gilmore12}), APOGEE (\citealt{Majewski16}) and the AMBRE Project (\citealt{deLaverny13}). All these observational data reveal a clear chemical distinction between a thick and a thin disc in the solar vicinity, with a dichotomy in $\alpha$-element\footnote{Elements characterised by capture of $\alpha$ particles. Examples are O, Mg, Si, S, Ca.} abundances (e.g. \citealt{Hayden14,Recio14,Rojas17,Mikolaitis17}).
Furthermore, GAIA data releases (DR1 and DR2, \citealt{Gaia16,Gaia18}) are even enhancing the values of these surveys, providing dynamical properties with unprecedent accuracy.

APOGEE data (\citealt{Hayden15}) like the other surveys reveal different abundance features for the two discs. However, at variance with other surveys, APOGEE spans an extended radial range, between 3 and 15 kpc from the Galactic centre. This allowed the community to study in detail the structure of the MW disc, and with a much larger statistical significance, due to the very large number of stars analysed ($\gtrsim$ 80 000).\\
In particular, the division of the disc in high-$\alpha$ (thick disc) and low-$\alpha$ (thin disc) sequences is confirmed at various radii. However, it is observed that the ratio between these populations changes with radius, with high-$\alpha$/low-$\alpha$ ratio progressively decreasing to outer radii (\citealt{Nidever14,Anders14}). At the same time, the scale heights $h_z$ of the two populations behave differently, with the low-$\alpha$ one showing a $h_z$ increasing with radius (\citealt{Bovy16}). These findings point towards a thick disc concentrated in the inner Galactic regions, with very short scale lenght (e.g. \citealt{Bovy16,Haywood16}) and not forming in an inside-out manner (\citealt{Haywood18}). \citet{Haywood16,Haywood18} and \citet{Miglio20} also determined a quenching period after the thick disc formation, however with uncertain origin (e.g. Galactic bar, see \citealt{Khoperskov18}). One of the explanations for this gap in the star formation at high redshift was the Gaia-Enceladus Sausage (GES, \citealt{Helmi18,Belokurov18,Koppelman19}) merger, where this major merger event (approximately 10 Gyr ago) heated up the gas in the dark matter halo (\citealt{Chaplin20}), preventing the SF in the disc (\citealt{Vincenzo19}). 

The hints coming from these surveys have suggested some revision on the "classical" MW chemical evolution scenarios (\citealt{Chiappini97,Chiappini01}).\\
As for example, \citet{Grisoni17} proposed a parallel scenario (see also \citealt{Chiappini09}), where MW thick and thin disc form in parallel but at different rates (see more recently \citealt{Grisoni19,Grisoni20}). In particular, this model allows to obtain a better agreement with AMBRE data than a two-infall model similar to that of \citet{Chiappini97,Romano10}, but revisited for the two discs only.\\
\citet{Noguchi18} and \citet{Spitoni19,Spitoni20} instead suggested two-infall models with a delayed formation of the thin disc. In this way \citet{Spitoni19,Spitoni20} were able to fit nicely not only the abundances, but also the stellar ages of the APOKASC sample (\citealt{Silva18}). \\
However, most of these works focus only on the solar vicinity, not accounting for the effects of their different prescriptions outside the "solar circle". An exception is represented by \citet{Grisoni18} where radial abundance gradients were treated, but mostly in the framework of one-infall models. \\

The aim of this paper is thus to study the effects of different prescriptions for MW chemical evolution models in a much more extended radial range, going from the most inner parts of the disc to Galactocentric radii much larger than solar ones.\\
To this aim, we consider MW two-infall models with "classical" prescriptions (from \citealt{Chiappini01,Colavitti09,Romano10,Grisoni17}) to reproduce present-day abundance and star formation rate (SFR) gradients, with or without detailed treatment of radial gas flows (\citealt{Spitoni11}).
The models which better reproduce the gradients are then compared with a large sample of stellar abundances at various Galactocentric distances, allowing changes in the model prescriptions to better explain the observational trends. We also explore the possible impact on MW gas accretion history of Gaia-Enceladus merger. To this aim, we run a detailed chemical evolution model that reproduce the abundances observed in the stellar relics of the merged galaxy following the work of \citet{Vincenzo19}. Following \citet{Vincenzo19}, we will refer to the Gaia-Enceladus merger scenario (\citealt{Helmi18}), where the stars of the accreted satellite are on retrograde orbits.

The paper is organised as follows. In Section \ref{s:data}, we show the observational data which have been considered. In Section \ref{s:models}, we describe the chemical evolution models adopted in this work. In Section \ref{s:results}, we present the comparison between model predictions and observations, discussing the implications of the obtained results. Finally, in Section \ref{s:conclusions}, we draw some conclusions.

\section{Abundance data}
\label{s:data}

In the first part of the paper, we concentrate on present-day radial gradients.
We use different tracers of abundance gradients at the present time, such as HII regions, Cepheids, young open clusters (YOC) and planetary nebulae (PNe) originating from young stars.

The HII regions data are taken from \citet{Esteban05,Rudolph06} and \citet{Balser15}. We are aware that potential biases between these abundance measurements can be present. In fact, to infer the abundances, different techniques are adopted (recombination excited lines (RELs) for \citealt{Esteban05}, collisionally excited lines (CELs) for \citealt{Rudolph06}, radio recombination lines (RRLs) for \citealt{Balser15}). However, the abundances are all inferred with the direct method, without the adoption of any strong-line metallicity calibration (e.g. $R_{23}$, $N_2$, $O_3N_2$; see \citealt{Pettini04}) that might have major impacts on the final values (\citealt{Kewley08,Bresolin16}).\\
PNe data are taken from \citet{Costa04} and \citet{Stanghellini18}. For the analysis of present time gradients, we consider only PNe whose progenitors are younger than 1 Gyr (YYPNe) from \citet{Stanghellini18}. At the same time, we take only objects classified as Type I (PNe with young progenitors) from \citet{Costa04}.\\
Regarding stellar data, the Cepheids abundances are taken from \citet{Luck11} and \citet{Genovali15}, while the adopted open cluster data are from \citet{Magrini17}. From the \citet{Magrini17} sample, we select the clusters with an age younger than 1.5 Gyr, in order to consider only YOC.

To give an estimate of the typical errors on the abundances, in the case of Cepheids they are of the order of $\sim$ 0.1 dex. Of the same order are typical abundance errors for PNe. For HII regions, typical errors are instead larger ($\sim$ 0.2 dex).\\
Regarding the errors on Galactocentric radius estimates, we note that the distance scale of PNe is much more uncertain than those of the other sources (\citealt{Grisoni18}). For Cepheids, YOC and HII regions errors in radius are generally of the order of $\lesssim 1$ kpc.

In plotting the Figures of present time abundance gradients, we also decide to bin the stellar and nebular (HII regions) data in radii bins of $2$ kpc of amplitude. To be more statistically consistent, within each bin we impose a minimum number of data of 10. In the case this number is not reached, the bin is enlarged until this condition is satisfied.\\

In the second part of the paper, we look at how the abundance pattern of [$\alpha$/Fe] vs. [Fe/H] varies with Galactocentric distance.\\
We adopt the APOGEE data of \citet{Hayden15}, who selected cool (3500 K$<T_{eff}<$5000 K) giant stars (1.0$<\log g <$3.8) with signal-to-noise ratio $(S/N)>$80 (see \citealt{Hayden15} for more details). All the stars are within 2 kpc of the midplane of the MW and have Galactocentric radii larger than 3 kpc.
Among the $\alpha$-elements, here we focus on [Mg/Fe] measurements, the most reliable from an observational point of view (see \citealt{Grisoni17,Grisoni18} and reference therein). 

Uncertainties on [X/H] abundance ratios are of $\sim$ 0.05 dex for both Fe and Mg (\citealt{Holtzman15}). On average, distances are accurate at the 15\%-20\% level (for a more detailed discussion on the distances see \citealt{Holtzman15}).

Due to the large amount of stellar data, for [Mg/Fe] vs. [Fe/H] diagrams we do not plot single data points but density regions of the data. This choice also allows us to better identify the stellar density in the abundance space and to compare it more clearly with model tracks.\\

In the discussion we also take advantage of stellar abundances combined with asteroseismic ages from the sample of \citet{Silva18}. The data are from an updated APOKASC (APOGEE+KeplerAsteroseismology  Science  Consortium, \citealt{Pinsonneault14}) catalogue, composed by 1197 stars selected from APOKASC (see \citealt{Spitoni19} for a more complete discussion on the observations and selection criteria).\\
In the sample that we use in this work we have not taken into account the so called "young $\alpha$ rich" (Y$\alpha$R) stars, whose origin is still not clear (e.g. \citealt{Chiappini15}).

Errors in stellar ages of \citet{Silva18} sample are strongly dependent on the age itself. In particular, the median uncertainty in the sample is 28.5\% of the stellar age. On the other hand, the errors on the stellar metallicity are independent from stellar ages and they are about 0.1 dex.

\section{Chemical evolution models}
\label{s:models}

\subsection{Milky Way two-infall models}
\label{ss:MW_model}

The framework of the MW chemical evolution models adopted in this paper is based on the two-infall model (e.g. \citealt{Chiappini97,Romano10}). In particular, we base on the revisited version of \citet{Grisoni17} for thick and thin discs only.\\
This model assumes that the MW forms as a results of two main infall episodes. The first episode forms the thick disc, whereas the second (delayed and slower) infall gives rise to the thin disc. In this model, the Galactic disc is approximated by rings 2 kpc wide. These rings can be either independent, without exchange of matter between them (e.g. \citealt{Chiappini97}), or can exchange gas due to radial flows (e.g. \citealt{Spitoni11}).\\

The basic equations that describe the evolution of a given chemical element $i$ are (see \citealt{Matteucci12} for the extended form):
\begin{equation}
\Dot{G}_i (R,t) = -\psi(R,t) X_i(R,t) + R_i(R,t) + \Dot{G}_{i,inf}(R,t)+\Dot{G}_{i,rf},
    \label{e:chemical_evo}
\end{equation}
where $G_i (R,t) = X_i (R,t) G(R,t)$ is the fraction of gas mass in the form of an element $i$  and $G(R,t)$ is the fractional mass of gas. The quantity $X_i(R,t)$ represents the abundance fraction in mass of a given element $i$, with the summation over all elements in the gas mixture being equal to unity.

The first term on the right hand side of Equation \eqref{e:chemical_evo} corresponds to the the rate at which an element $i$ is removed from the ISM due to the star formation process. The SFR is parametrised according to the Schmidt-Kennicutt law (\citealt{Kennicutt98}):
\begin{equation}
\psi(R,t) = \nu \Sigma_{gas}(R,t)^k,
    \label{e:SFR}
\end{equation}
where $\Sigma_{gas}$ is the surface gas density, $k=1.5$  is  the law index  and $\nu$ is the star formation  efficiency.

$R_{i}(R,t)$ (see \citealt{Palla20} for the complete expression) takes into account the nucleosynthesis from low-intermediate mass stars (LIMS, $m < 8$M$_\odot$), core collapse (CC) SNe (Type II and Ib/c, $m > 8$M$_\odot$) and Type Ia SNe. For these latter, we assume the single-degenerate (SD) scenario, in which a C-O white dwarf in a binary system accretes mass from a non-degenerate companion until it  reaches nearly the Chandrasekhar mass ($\sim 1.44$M$_\odot$).
As in many previous two-infall model papers (e.g. \citealt{Romano10,Grisoni17,Spitoni19}), we adopt the SD delay-time-distribution (DTD) function from \citet{MatteucciRecchi01}. In this formalism, the clock for the explosion is given by the lifetime of the secondary star. In \citet{Matteucci09} it has been shown that this DTD and its resulting abundance patterns are very similar to those of the double degenerate (DD) model for Type Ia SNe of \citet{Greggio05}.\\
The stellar yields are taken from \citet{Karakas10} (LIMS), \citet{Kobayashi06} (CC-SNe) and \citet{Iwamoto99} (Type Ia SNe). The fraction of stars in binary systems able to originate Type Ia SNe is fixed at a value able to reproduce the present-day Type Ia SN rate (0.035, see \citealt{Romano05}).\\
We note that $R_i(R,t)$ depends also on the initial mass function (IMF). The adopted IMF is the \citet{Kroupa93} one.

The  third  term  in  Equation \eqref{e:chemical_evo}  is  the gas infall rate. In  particular,  in  the  two-infall  model the gas accretion  is computed in this way:
\begin{multline}
    \Dot{G}_{i,inf}(R,t)=A(R)X_{i,1inf}(R)e^{-\frac{t}{\tau_1}} +\\ +\theta(t-t_{max}) B(R) X_{i,2inf}(R) e^{-\frac{t-t_{max}}{\tau_2(R)}},
    \label{e:infall}    
\end{multline}
where $G_{i,inf}(R,t)$ is the infalling material in the form of element $i$ and $X_{i,Qinf}$ is the composition of the infalling gas for the $Q$th infall. In this work, we try both primordial and pre-enriched chemical compositions for the two infall episodes. We remind the reader that the $\theta$ in the Equation above is the Heavyside step function.\\
$t_{max}$ is the time for the maximum mass accretion onto the Galactic disc and corresponds to the start of the second infall phase: we let this parameter vary in the paper in order to explain the observational data.  $\tau_1$ and $\tau_2(R)$ represent the infall timescales for the thick and thin discs, respectively. In this paper, we assume that the timescale for mass accretion in the Galactic thin disc increases linearly with radius according to the inside-out scenario (e.g. \citealt{Chiappini01,Cescutti07}):
\begin{equation}
    \tau_2(R) = \bigg( 1.033\, \frac{R}{kpc} - 1.267\bigg) Gyr.
    \label{e:inside_out}    
\end{equation}
We perform also calculations with other inside-out laws, either flatter or steeper:
\begin{equation*}
    \tau_2(R)=\bigg([0.75,1.25] \frac{R}{kpc} + [0.997,-3.003] \bigg) Gyr,
\end{equation*}
(in brackets are the ranges of parameters explored) finding no substantial effects on the results. For this reason, in this paper we only show models adopting Equation \eqref{e:inside_out}. We highlight that $\tau_2(8$kpc$)\simeq 7$ Gyr is fixed by reproducing the G-dwarf metallicity distribution in the solar vicinity.
The timescale of the thick disc is instead fixed to 1 Gyr (see also \citealt{Haywood18}).\\
The quantities $A(R)$ and $B(R)$ are two parameters fixed by  reproducing  the  present  time  total surface  mass density for the thick and thin discs at a radius $R$. In this paper, we assume that the surface mass density in the Galactic thin disc follows an exponential profile:
\begin{equation}
    \Sigma(R)=\Sigma_0 e^{-R/R_d},
    \label{e:Sigma}
\end{equation}
where $\Sigma_0=531$ M$_\odot$ pc$^{-2}$ is the central surface mass density and $R_d=3.5$ kpc is the disc scale length (see e.g., \citealt{Spitoni17}). For the thick disc we explore different kind of surface profiles (inverse linear, exponentially decaying), fixing only the density in the solar neighborhood ($R =$8 kpc) at $\sim12$ M$_\odot$ pc$^{-2}$. These parameterisations for the discs allow us to obtain: (i) a $\Sigma_{thin}(8$kpc$)\sim54$ M$_\odot$ pc$^{-2}$, which is in agreement with the values for thin disc given by e.g., \citet{Bovy13,Read14}, and (ii) a ratio $\Sigma_{thin}$(8 kpc)$/\Sigma_{thick}$(8 kpc)$\sim 4$, in agreement with recent chemical evolution works on the solar neighbourhood (\citealt{Spitoni20}).\\

As a summary,  the physical parameters which have been left constant in all models are listed in Table \ref{t:fixed_param}.

\begin{table}
    \centering
    \caption{Fixed parameters for two-infall chemical evolution models adopted in this work. In the first column we list the variables that remain constant. In the last column, we show the adopted parameterization.}
    \begin{tabular}{c | c }
        \hline
        \hline
         Variable & Fixed parametrization \\[0.05cm]
         \hline
         $\nu_{thick}$ & $2$ Gyr$^{-1}$\\[0.15cm]
         $\tau_1$ & $1$ Gyr\\[0.15cm]
         $\tau_2(R)$ & $(1.033R-1.267)$ Gyr\\[0.15cm]
         $\Sigma_{thin}(R)$ & $531 e^{-R/3.5}$ M$_\odot$ pc$^{-2}$\\[0.15cm]
         $\Sigma_{thick}(R=8)$ & $12.3$ M$_\odot$ pc$^{-2}$ \\
         \hline
         \hline
    \end{tabular}\\
    \label{t:fixed_param}
    \textbf{Note} $R$ is expressed in kpc.
\end{table}

\subsubsection{Implementation of radial inflows}
\label{sss:radial_flows}
The last term of Equation \eqref{e:chemical_evo} refers to radial inflows of gas. We implement them following the prescriptions of \citet{Portinari00,Spitoni11}.

We  define  the $k_{th}$  shell  in  terms  of  the  Galactocentric  radius $R_k$, its inner and outer edges being labelled as $R_{k-1/2}$ and $R_{k+1/2}$. Through these edges, gas inflow occurs with velocities $\upsilon_{k-1/2}$ and $\upsilon_{k+1/2}$, respectively; the flow velocities are assumed to be positive outwards and negative inwards.\\
Radial inflows with a flux $F(R)$, alter the gas surface density in the $k_{th}$ shell according to
\begin{equation}
    \Dot{\Sigma}_{gas, rf}(r_k,t)=-\frac{1}{\pi(R^2_{k+1/2}-R^2_{k-1/2})}[F(R_{k+1/2})-F(R_{k-1/2})],
    \label{e:flow_explain}
\end{equation}
where
\begin{equation*}
    \begin{array}{l}
         F(R_{k+1/2}) = 2\pi R_{k+1/2} \upsilon_{k+1/2} \Sigma_{gas}(R_{k+1},t),  \\[0.1cm]
         F(R_{k-1/2}) = 2\pi R_{k-1/2} \upsilon_{k-1/2} \Sigma_{gas}(R_{k},t).
    \end{array}
\end{equation*}
Since the edges $R_{k-1/2}$ and $R_{k+1/2}$ are taken respectively at the mid-point between the shells $k$, $k-1$ and $k$, $k+1$, we have that:
\begin{equation*}
    (R^2_{k+1/2}-R^2_{k-1/2})= \frac{R_{k+1}-R_{k-1}}{2} \bigg(R_k + \frac{R_{k-1}+R_{k+1}}{2} \bigg).
\end{equation*}
Inserting the latter expressions into Equation \eqref{e:flow_explain} we finally obtain:
\begin{equation}
    \Dot{G}_{i,rf}(R_k,t)=-\beta_k G_i(R_k,t) + \gamma_k G_i(R_{k+1},t),
    \label{e:radial_flows}
\end{equation}
where
\begin{equation*}
    \begin{array}{l}
    \beta_k=-\frac{2}{R_k+\frac{R_{k-1}+R_{k+1}}{2}} \big[ \upsilon_{k-1/2} \frac{R_{k-1}+R_k}{R_{k+1}-R_{k-1}} \big],\\[0.2cm]
    \gamma_k=-\frac{2}{R_k+\frac{R_{k-1}+R_{k+1}}{2}} \big[ \upsilon_{k+1/2} \frac{R_{k}+R_{k+1}}{R_{k+1}-R_{k-1}} \big] \frac{\Sigma_{k+1}}{\Sigma_{k}}.
    \end{array}
\end{equation*}
Here, $\Sigma_{k}$ and $\Sigma_{k+1}$ are the present total surface mass density at the radius $R_k$ and $R_{k+1}$, respectively. In our formulation, we assume that there are no flows from the outer parts of the disc where there is no star formation: flows start at radius $18$ kpc and move inward.

In our models, we test different speed patterns. In particular, we see the effects of radial flows with fixed velocity $\upsilon_{flow}$ ($-1$ km s$^{-1}$, e.g. \citealt{Spitoni11}), radius dependent velocity ($\upsilon_{flow}=-\big(\frac{R}{4}-1\big)$, e.g. \citealt{Grisoni18}) and both radius and time dependent speed.\\
The latter formulation represents a novelty relative to previous \citet{Portinari00,Spitoni11} radial gas flows implementations.  We adopt a velocity that in the thin disc phase decreases with time, following the exponential behaviour of the second infall ($e^{-\frac{t-t_{max}}{\tau_2}}$). A flow speed that follows the infall is a reasonable assumption: the ratio infall rate/gas density, which drives the variations in the flow speed (Equation 28 of \citealt{Pezzulli16}, see also \citealt{Lacey85}) decrease with time and radius, as the infall rate in an inside-out scheme.
A decrease of the flow velocity with time is also supported by \citet{Bilitewski12} who, using a different flow implementation, computed radial flows speed patterns by imposing the conservation of angular momentum. They also found that the  inflow speed generally decreases with decreasing radius, as done in this work.\\
Concerning the velocity range of all the radial flow prescriptions, they span the interval $0-\sim4$ km s$^{-1}$. This is in agreement with previous chemical evolution studies (e.g. \citealt{Bilitewski12,Mott13,Grisoni18,Vincenzo20}) as well as observations of external galaxies (e.g. \citealt{Wong04}).

\subsection{Enceladus model}
\label{ss:Enceladus_model}

The recent discovery of Gaia-Enceladus (\citealt{Helmi18,Belokurov18}), suggested that the inner Galactic halo and the thick disc might have formed by means of an important accretion episode  involving a mass similar to that of the Small Magellanic Cloud, which occurred roughly 10-11 Gyr ago (\citealt{Chaplin20}). These findings (supported by GAIA DR2 data) confirmed the long standing suggestion of a significant merger occurred 10-12Gyr, which was concomitant with the formation of the present-day thick disc (see also \citealt{Wise01} for a review).\\
Here we run a model for Enceladus itself by assuming the same parameters as in \citet{Vincenzo19} and then we simulate the formation of the MW discs by gas accretion episodes with the same chemical composition as predicted by our model and in agreement with the measured abundances in Gaia-Enceladus. We discuss various cases where the thick or the thin disc are formed by enriched infall including also the chemical composition of Enceladus.

\begin{table}
    \centering
    \caption{Input parameters for Enceladus chemical evolution model. In the first column we list the input variables of the model. In the second column, we show the adopted values.}
    \begin{tabular}{c | c }
        \hline
        \hline
         Input parameter & Adopted value \\[0.05cm]
         \hline
         $M_{inf,Enc}$ & $10^{10}$ M$_\odot$\\[0.15cm]
         $\tau_{inf,Enc}$ & $0.24$ Gyr\\[0.15cm]
         $\nu_{Enc}$ & $0.42$ Gyr$^{-1}$\\[0.15cm]
         $\omega_{Enc}$ & $0.5$\\
         \hline
         \hline
    \end{tabular}\\
    \label{t:Enceladus_param}
    We adopt a \citet{Kroupa93} IMF.\\ Yields are from \citet{Karakas10} (LIMS), \citet{Kobayashi06} (CC-SNe) and \citet{Iwamoto99} (Type Ia SNe).
\end{table}

\begin{table*}
    \centering
    \caption{Variable parameters for two-infall chemical evolution models adopted in this work. In the first column, we write the name of the model. In the second column, there is the density profile for the thick disc $\Sigma_{thick}(R)$. In the third column, we indicate the time of maximum infall on the thin
disk $t_{max}$. In the fourth column we show the level of enrichment of the second infall episode. In the fifth column, we list the star formation efficiency of the thin disc $\nu_{thin}$ at different radii (4-6-8-10-12-14-16-18 kpc) from the Galactic center. In the last column, we indicate the radial flow speed pattern (where present).
    The horizontal line divides the models with "classical" two-infall model features ($t_{max}$, $\Sigma_{thick}(R)$, infall [Fe/H] as in \citealt{Chiappini01,Colavitti09,Romano10}) from the ones with revised values.} 
    \begin{tabular}{c | c  c  c  c  c}
    \hline
    \hline
    Model & $\Sigma_{thick}(R)$ & $t_{max}$ & Infall [Fe/H] & $\nu_{thin}$ (4-6-8-10-12-15-16-18 kpc) & Radial flows \\[0.05cm]
    & [M$_\odot$ pc$^{-2}$] & [Gyr] & [dex] & [Gyr$^{-1}$]  & [km s$^{-1}$]\\
    \hline
    MW A & $\propto\frac{1}{R}$ & 1 & primordial & 1 & -\\[0.15cm]
    MW B & $\propto\frac{1}{R}$ & 1 & primordial & 5, 2, 1, 0.75, 0.5, 0.25, 0.15, 0.1 & -\\[0.15cm]
    MW C & $\propto\frac{1}{R}$ & 1 & primordial & 1 & $-1$\\[0.15cm]
    MW D &
    
    $\propto\frac{1}{R}$ & 1 & primordial & 1 & $-\upsilon(R)=-\big(\frac{R}{4}-1\big)$ if $t>t_{max}$; $-1$ if $t<t_{max}$\\[0.15cm]
    MW E & $\propto\frac{1}{R}$ & 1 & primordial & 1 & $-\big[1+\upsilon(R)\cdot e^{-\frac{t-t_{max}}{\tau_2}}\big]$ if $t>t_{max}$; $-1$ if $t<t_{max}$\\[0.15cm]
    MW F & $\propto\frac{1}{R}$ & 1 & primordial & 5, 2, 1, 0.75, 0.5, 0.25, 0.15, 0.1 & $-1$\\[0.15cm]
    MW G & $\propto\frac{1}{R}$ & 1 & primordial & 5, 2, 1, 0.75, 0.5, 0.25, 0.15, 0.1 & $-\upsilon(R)$ if $t>t_{max}$; $-1$ if $t<t_{max}$\\[0.15cm]
    \hline
    MW E1 & $\propto\frac{1}{R}$ & 3.25 & primordial & 1 & $-\big[1+\upsilon(R)\cdot e^{-\frac{t-t_{max}}{\tau_2}}\big]$ if $t>t_{max}$; $-1$ if $t<t_{max}$\\[0.15cm]
    MW F1 & $\propto\frac{1}{R}$ & 3.25 & primordial & 5, 2, 1, 0.75, 0.5, 0.25, 0.15, 0.1 & $-1$\\[0.15cm]
    MW E2 & $\propto\frac{1}{R}$ & 4.5 & primordial & 1 & $-\big[1+\upsilon(R)\cdot e^{-\frac{t-t_{max}}{\tau_2}}\big]$ if $t>t_{max}$; $-1$ if $t<t_{max}$\\[0.15cm]
    MW F2 & $\propto \frac{1}{R}$ & 4.5 & primordial & 5, 2, 1, 0.75, 0.5, 0.25, 0.15, 0.1 & $-1$\\[0.15cm]
    MW E3 & $\propto e^{-R/4}$ & 3.25 & primordial & 1 & $-\big[1+\upsilon(R)\cdot e^{-\frac{t-t_{max}}{\tau_2}}\big]$ if $t>t_{max}$; $-1$ if $t<t_{max}$\\[0.15cm]
    MW F3 & $\propto e^{-R/4}$ & 3.25 & primordial & 5, 2, 1, 0.75, 0.5, 0.25, 0.15, 0.1 & $-1$\\[0.15cm]
    MW E4 & $\propto e^{-R/2.3}$ & 3.25 & primordial & 1 & $-\big[1+\upsilon(R)\cdot e^{-\frac{t-t_{max}}{\tau_2}}\big]$ if $t>t_{max}$; $-1$ if $t<t_{max}$\\[0.15cm]
    MW F4 & $\propto e^{-R/2.3}$ & 3.25 & primordial & 5, 2, 1, 0.75, 0.5, 0.25, 0.15, 0.1 & $-1$\\[0.15cm]
    MW E5 & $\propto e^{-R/2.3}$ & 3.25 & -1 ($R<8$) & 1 & $-\big[1+\upsilon(R)\cdot e^{-\frac{t-t_{max}}{\tau_2}}\big]$ if $t>t_{max}$; $-1$ if $t<t_{max}$\\[0.15cm]
    MW F5 & $\propto e^{-R/2.3}$ & 3.25 & -1 ($R<8$) & 5, 2, 1, 0.75, 0.5, 0.25, 0.15, 0.1 & $-1$\\[0.15cm]
    MW E6 & $\propto e^{-R/2.3}$ & 3.25 & -0.5 ($R<8$) & 1 & $-\big[1+\upsilon(R)\cdot e^{-\frac{t-t_{max}}{\tau_2}}\big]$ if $t>t_{max}$; $-1$ if $t<t_{max}$\\[0.15cm]
    MW F6 & $\propto e^{-R/2.3}$ & 3.25 & -0.5 ($R<8$) & 5, 2, 1, 0.75, 0.5, 0.25, 0.15, 0.1 & $-1$\\[0.15cm]
    MW E7 & $\propto e^{-R/2.3}$ & 3.25 & 0 ($R<8$) & 1 & $-\big[1+\upsilon(R)\cdot e^{-\frac{t-t_{max}}{\tau_2}}\big]$ if $t>t_{max}$; $-1$ if $t<t_{max}$\\[0.15cm]
    MW F7 & $\propto e^{-R/2.3}$ & 3.25 & 0 ($R<8$) & 5, 2, 1, 0.75, 0.5, 0.25, 0.15, 0.1 & $-1$\\[0.15cm]
    \hline
    \hline
\end{tabular}
\\[0.1cm]
\textbf{Notes} $R$ is expressed in kpc, $t$ is expressed in Gyr. The minus sign in the last column means that the flow is an inflow. 
To better identify the models, we distinguish them with letters if they adopt the same $t_{max}$, $\Sigma_{thick}(R)$, infall [Fe/H] and with numbers if at least one of these latter parameters is changed.
\label{t:models}
\end{table*}

In \citet{Vincenzo19}, a model characterised by a small star formation efficiency, fast infall time-scale and a mild outflow could nicely reproduce the observed [$\alpha$/Fe] ratios as well as the metallicity distribution function (MDF) measured in Enceladus. They also predicted a stellar mass for the galaxy of $\sim 5 \cdot 10^{9}M_{\odot}$, and suggested that the merging event could have been the cause for halting the star formation in the MW 10 Gyr ago, a fact already predicted by chemical models (e.g. \citealt{Noguchi18}).\\ 
The input parameters for Enceladus model are listed in Table \ref{t:Enceladus_param}, together with the adopted IMF and stellar yields. We note in the last row the parameter $\omega$: this is the wind mass loading factor, a free parameter of the model which regulates the gas outflow rate. In fact, the outflow rate is expressed as $\dot{G}_{out}(t)=\omega\,\psi(t)$, where $\psi(t)$ is the SFR.


\begin{figure*}
    \centering
    \includegraphics[width=.9\textwidth]{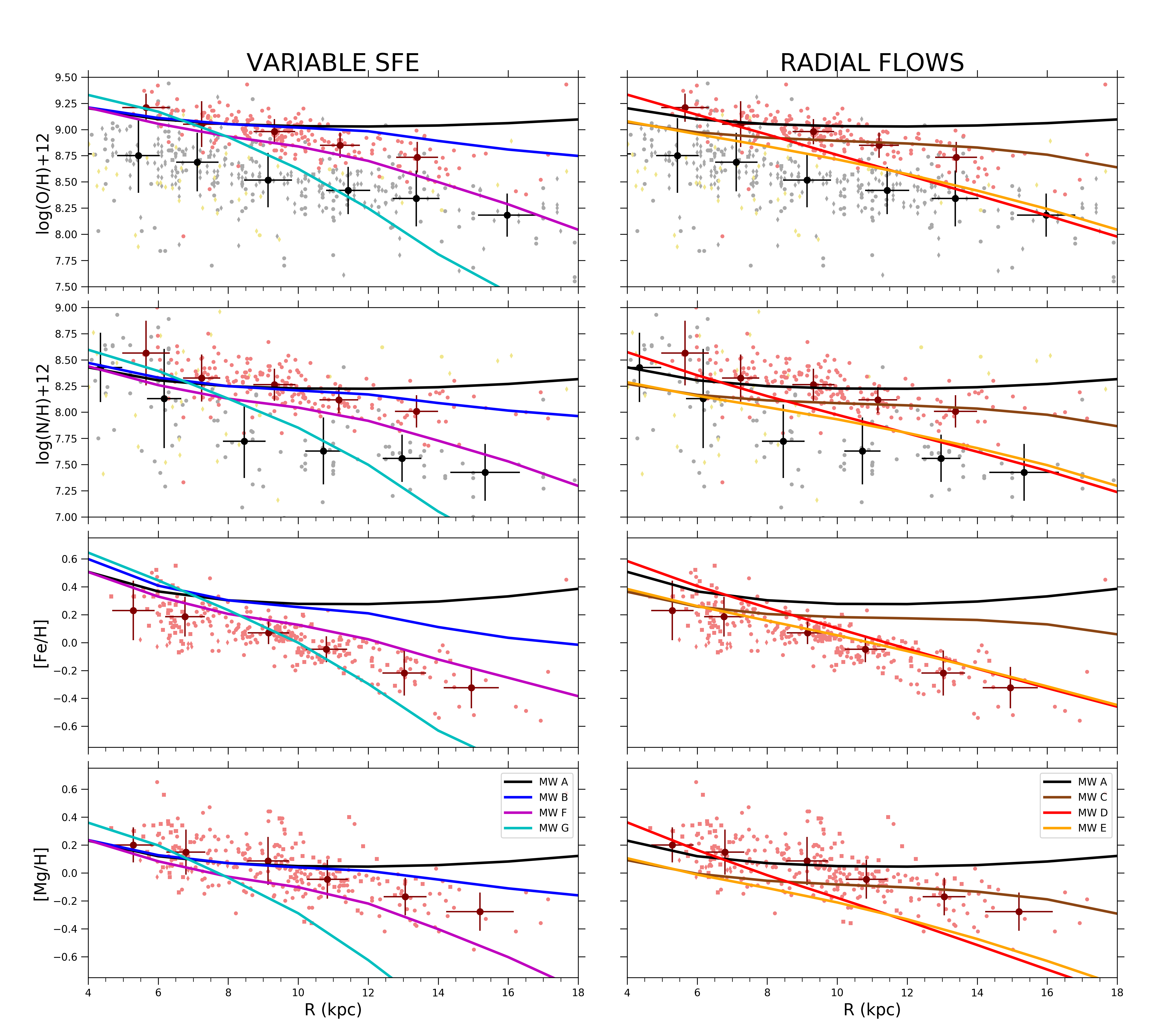}
    \caption{Observed and predicted radial abundance gradients for models that adopt a variable SFE with radius (left panels) and models that adopt only radial flows (right panels). Abundances considered are oxygen (first row), nitrogen (second row), iron (third row) and magnesium (lower row). Data are from \citet{Esteban05} (grey squares), \citet{Rudolph06} (grey dots) and \citet{Balser15} (grey diamonds) for HII regions, \citet{Luck11} (light red dots) and \citet{Genovali15} (light red squares) for Cepheids, \citet{Magrini17} (light red diamonds) for YOC, \citet{Costa04} (khaki dots) and \citet{Stanghellini18} (khaki diamonds) for PNe. Black and dark red points with errorbars are data bins with associated rms for HII regions and stellar data, respectively.}
    \label{f:gradients_present}
\end{figure*}


\section{Results}
\label{s:results}


We run several models, keeping fixed the IMF, the efficiency of star formation in the thick disc, the timescales for the formation of the thick and thin discs, the present time distribution of the surface mass density along the thin disc and the present time surface density for the thick disc in the solar neighbourhood (see Table \ref{t:fixed_param}).\\ 
In Table \ref{t:models} we show the different parameters adopted by the different models. In particular, we divide Table \ref{t:models} in two parts: the upper part refers to the models adopting "classical" two-infall prescriptions, i.e. $t_{max}$, $\Sigma_{thick}(R)$, second infall [Fe/H], as in \citet{Chiappini01,Colavitti09,Romano10,Grisoni17}. In the lower part, instead, we list the models with revised values of the above parameters. 
To better identify the models, we distinguish them with letters (from A to G) if they adopt the same $t_{max}$, $\Sigma_{thick}(R)$, infall chemical composition (indicated by [Fe/H]) and with numbers if at least one of these latter parameters is changed.

\subsection{Present-day gradients}
\label{ss:presentday_results}

We first focus on the analysis of present time abundance gradients along the thin disc. We look in particular at the different results given by the models of Table \ref{t:models} with "classical" two-infall model prescriptions ($t_{max}=1$ Gyr, $\Sigma_{thick}\propto 1/R$: Models MW A to G). For the analysis of present-day abundances we choose to show the abundances of oxygen, nitrogen, magnesium and iron. In this way, we analyse elements produced on different timescales and with different origin (primary or secondary origin\footnote{Primary production of an element stems directly from H and He synthesis. In the case of secondary production, the seed for the synthesis must be a metal (e.g. C, O)}).\\

Looking at Figure \ref{f:gradients_present}, we note the division between left and right panels. On the left side, we show models that adopt different SFEs at different radii (B, F and G). On the right side, we show the models which adopt only radial inflows (C, D and E). In addition, in both sides, we show in black a standard model (Model A) with constant SFE, no radial gas flows and inside-out formation for the thin disc ($\tau_2(R) \propto R$) by means of primordial gas. The model results are compared with the adopted data sets (i.e. HII regions, Cepheids).\\
As it can be seen from the first and second row panels, there is an offset between nebular (HII regions and PNe) and stellar (Cepheids and YOC) data. This offset can be interpreted as a real observational bias in at least one of the two observational techniques. In fact, nitrogen is minimally depleted by dust in the MW ([N/H]$_{ISM}$-[N/H]$_{gas}<$0.1 dex, see \citealt{Jenkins09}). In this way, the offset between nebular and stellar data for N and O can be interpreted in terms of a bias between the two observational techniques.\\
Still concerning nitrogen, we signal that adopting stellar  yields with primary N production by all massive stars of all metallicities\\ (\citealt{Matteucci86}) does not change significantly the model results on present-day abundances.

\begin{table*}
    \centering
    \caption{Present-day slopes for elemental abundance gradients from the best models of Figure \ref{f:gradients_present} (MW B, MW F, MW D, MW E) and observed slopes from HII regions, Cepheids and YOC. In the first column, we write the elements considered in the analysis. From second to fifth column we list the results obtained by chemical evolution models. In the last two columns, we indicate the results coming from observations.} 
    \begin{tabular}{c | c  c  c  c | c  c }
    \hline
    \hline
     & \multicolumn{4}{c}{CHEMICAL EVOLUTION MODELS (4-16 kpc)} & \multicolumn{2}{c}{OBSERVATIONS} \\[0.1cm]
    $d$(X/H)$/dR$ & MWD B & MW F & MW D & MW E & HII regions & Cepheids+YOC \\
    & (dex$/$kpc) & (dex$/$kpc) & (dex$/$kpc) &	(dex$/$kpc) &	(dex$/$kpc) & (dex$/$kpc) \\
    \hline
    O & -0.0305  & -0.0734 & -0.0961 & -0.0688 &  -0.0539 & - 0.0497\\
    N & -0.0344 & -0.0713 & -0.0931 & -0.0646 &  -0.0888 & -0.0423\\
    Fe & -0.0424 & -0.0599 & -0.0751  & -0.0571 &  - & -0.0556\\
    Mg & -0.0237 & -0.0656 & -0.0865 & -0.0598 &  - & -0.0438\\
    \hline
    \hline
\end{tabular}
\label{t:slopes}
\end{table*}

In Figure \ref{f:gradients_present} we can see that only a few models can reasonably  reproduce the abundance gradients. In particular, the Model MW A with only inside-out for the formation of the thin disc does not reproduce the observed gradients, since it predicts a very flat behaviour;  even adopting more extreme inside-out laws (see Section \ref{ss:MW_model}) the problem remains. On the other side, models with either variable efficiency of star formation or radial gas flows or both produce results in agreement with observations. In fact, a SFE decreasing with radius boosts the metal enrichment in the inner regions relative to the outer ones. Radial flows, instead, bring the material from the outer regions to the inner ones, providing an additional source of gas to enhance the SF (and so the metal enrichment).\\
Among the models of the left panel of Figure \ref{f:gradients_present} Model MW F seems to be the best one, while Model MW E seems to be the best among those of the right panel. Model MW F assumes a variable efficiency of star formation along the disc plus radial gas flows with constant speed, whereas Model MW E assumes a constant efficiency of star formation but radial gas flows with speed variable with radius and time (see Section \ref{sss:radial_flows}). These results are supported by previous findings (see for example \citealt{Colavitti09,Grisoni18}) suggesting that either variable SFE or radial gas flows do reproduce the gradients along the thin disc of the Milky Way.\\
Looking at the other models, on the left side we see that the model with variable SFE only (MW B) predicts a steeper slope than the MW A model, in acceptable agreement with data from Cepheids and YOC (except for [Fe/H]). The agreement is absent instead for model MW G (variable SFE + $\upsilon_{flow}=\big(R/4-1\big)$ km s$^{-1}$), which shows too steep gradients to be consistent with the observations. On the right side, we note that the model with constant inflow speed (MW C) predicts too flat gradients to agree with the adopted data sets. MW D model instead predicts relatively too steep slopes in the inner regions of the Galaxy, and values very similar to model MW E in the outer regions.

\begin{figure}
    \centering
    \includegraphics[width=1.\columnwidth]{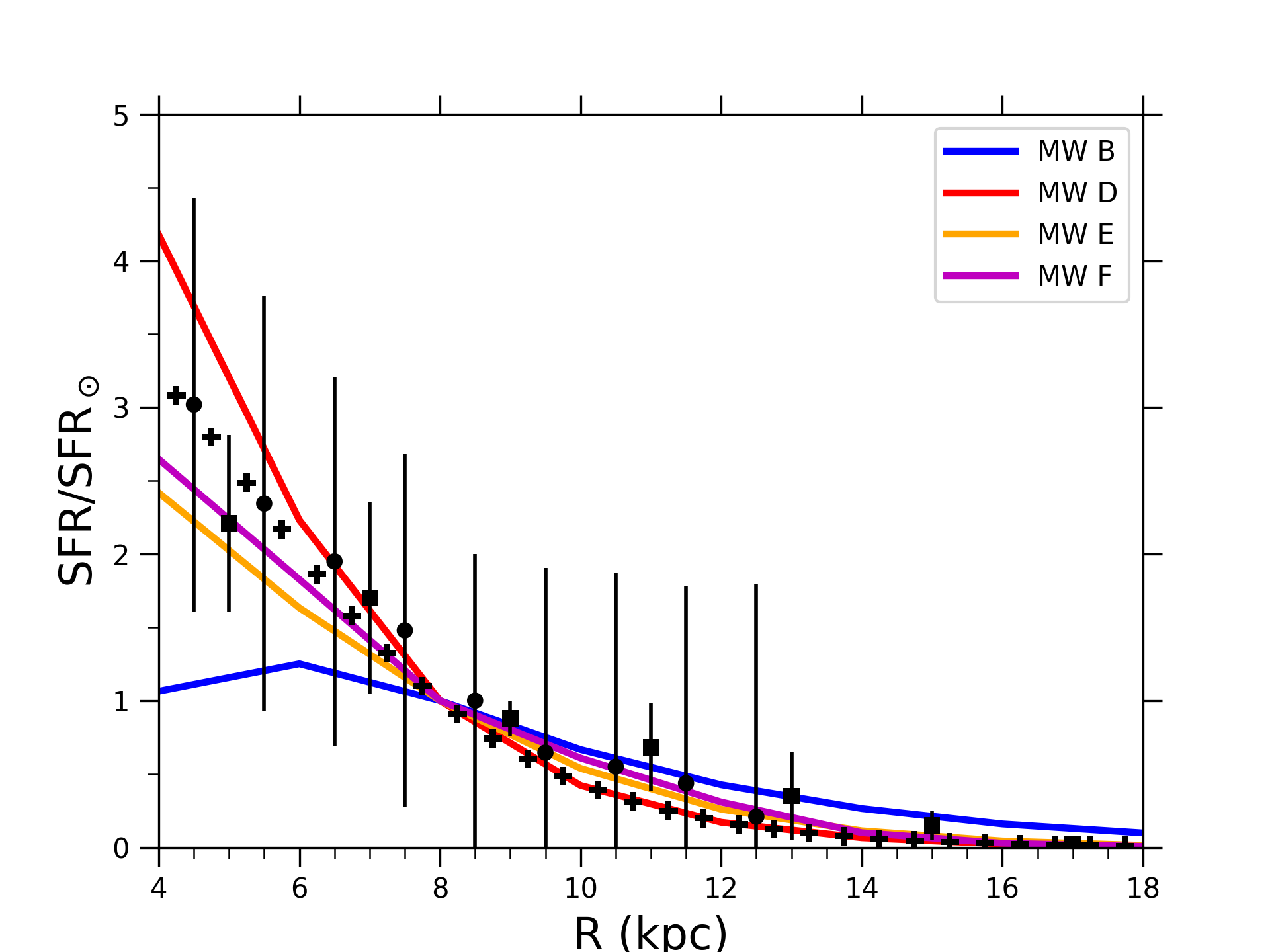}
    \caption{Observed and predicted radial SFR density gradient relative to the solar neighbourhood. Lines are computed for the MW B model (solid blue), MW F (solid magenta), MW D (solid red) and MW E (solid orange). Data with errorbars are from \citet{Rana91} (black points) and \citet{Stahler05} (black squares). Black crosses with no error bars follow the analytical form suggested by \citet{Green14} for the MW SFR profile.}
    \label{f:SFR_present}
\end{figure}

In Table \ref{t:slopes} we show the slopes obtained by means of the models in better agreement with observational data. In particular, we decide to compare the fits of the data with the ones of the models in the region 4-16 kpc. This choice is justified by the very limited amount of nebular and stellar data in the outer regions ($R\gtrsim$ 16 kpc).\\
As we can see from Table \ref{t:slopes}, the values predicted by the models with radius and time dependent radial flows (MW E) and combined variable SFE and radial flows (MW F) best agree with the fits to the data. The other two models in the Table \ref{t:slopes} show larger (MW D) and smaller (MW B) negative gradients, respectively. Concerning MW D model, we have only marginal agreement with the N/H gradient slope from HII regions data. However, we take this data set with caution, because of the low number statistic together with the quite large observed data spread. We also note that the gradient  predicted by model MW B is flatter than those obtained by other studies that adopt a variable SFE (e.g. \citealt{Grisoni18}): this is due to the milder SFE-radius relation adopted here (see Table \ref{t:models}). In principle, the adoption of a steeper SFE gradient can produce a better agreement with the observed abundances: however, we will see later that the SFR-radius relation does not favour such an assumption (see Figure \ref{f:SFR_present}). \\

Most of the models of Table \ref{t:slopes} (MW D, E and F) do also well reproduce the SFR gradient along the thin disc, as shown in Figure \ref{f:SFR_present}. In fact, in Figure \ref{f:SFR_present} we compare the present-day SFR gradient (relative to the value in the solar neighbourhood) of the 4 models shown in Table \ref{t:slopes} (MW B, D, E, F) with available data from the literature.

\begin{figure}
    \centering
    \includegraphics[width=1.\columnwidth]{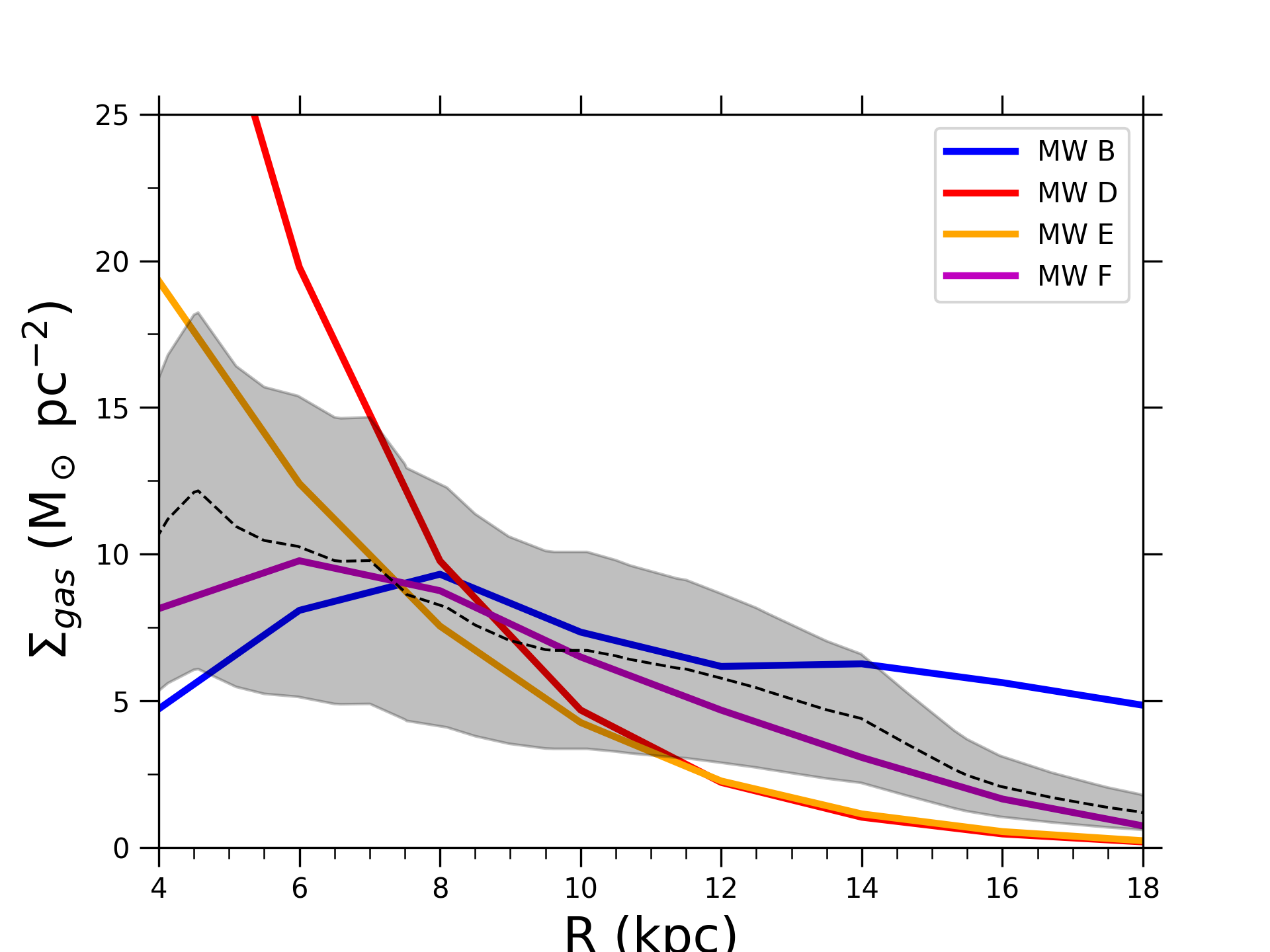}
    \caption{Observed and predicted radial gas surface density gradient. Lines are computed for the MW B model (solid blue), MW F (solid magenta), MW D (solid red) and MW E (solid orange). The black dashed line is the average value between \citet{Dame93} and \citet{Nakanishi03,Nakanishi06} datasets. The shaded region represents the typical uncertainty at each radius, for which we adopt either 50\% of the average (see \citealt{Nakanishi06}) or half the difference between the minimum and maximum values in each radial bin (if larger).}
    \label{f:gas_present}
\end{figure}

A general decrease of the SFR with Galactocentric radius is seen for all the models, in agreement with observations.\\
However, Model MW B (variable SFE only) fails to reproduce the observed behaviour in the inner regions. This finding also excludes the adoption of steeper SFE-radius relations, which instead can give better agreement with abundance gradient data. In fact, larger SFEs produce lower gas consumption timescales, which tend to lower the SFR at late times. \\
All the other models show a good agreement with \citet{Rana91} data set, which has however very large error bars. For this reason, the values taken from \citet{Stahler05} and \citet{Green14} are more indicative. The \citet{Stahler05} data set is formed by weighted mean values (on individual data errors) of SFR from SN remnants, pulsars and HII regions. The comparison with models shows that MW F and MW E models are the most consistent with the observed trend, in particular in the inner Galactic regions. For what concerns the \citet{Green14} relation, it is a modified version of \citet{Lorimer06} analytical fit to the pulsar distribution:
\begin{equation}
    SFR(R)/SFR_\odot= \bigg(\frac{R}{R_0}\bigg)^{b}\, e^{-c\big(\frac{R-R_0}{R_0}\big)},
\end{equation}
where $R_0=8$ kpc and parameters $b=2$ and $c=5.1$ (in \citealt{Lorimer06} $b=1.9$ and $c=5$).
The fit is in very good agreement with SN remnants compilation by \citet{Green14}, as well as for \citet{Urquhart14} data on luminous massive stars (although with larger scatter). As can be seen in Figure \ref{f:SFR_present}, the relation is very nicely reproduced by model MW D up to the solar radius, whereas stay in between MW F and MW D models in the inner Galactic radii.\\


\begin{figure*}
    \centering
    \includegraphics[width=1\textwidth]{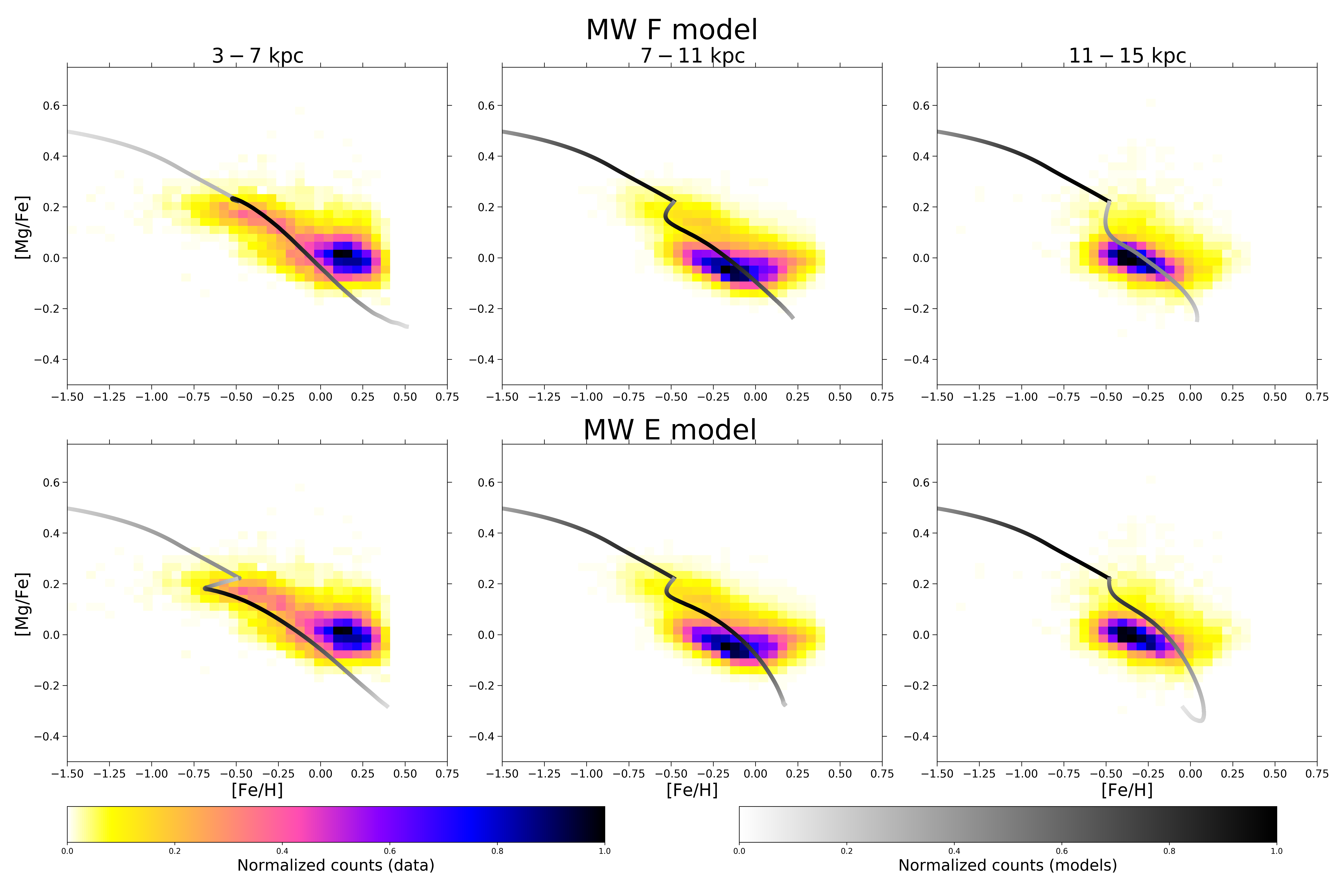}
    \caption{[$\alpha$/Fe] vs. [Fe/H] abundance ratios of model MW F (upper panel) and MW E (lower panel), with "classical" two-infall prescriptions ($t_{max}=1$ Gyr, $\Sigma_{thick}\propto1/R$).
    Data are from \citet{Hayden15}. Left, central and right panels show the models and data in the ranges 3 < $R/$kpc <7, 7 < $R/$kpc < 11 and 11 < $R/$kpc < 15, respectively.
    Left colorbar indicates the normalised counts of data in a certain bin of the plot (the area of each bin is fixed at the value of 0.056 dex$\times$0.031 dex), while right colorbar indicates the normalised counts of stars formed in the model at a certain time t (and hence at certain [Mg/Fe] and [Fe/H]).}
    \label{f:MgFe_bestgradient}
\end{figure*}

We also analyse the predicted profiles of the gas along the Galactic disc according to the different prescriptions adopted in this work. In particular, as for the SFR, we concentrate on the models showing a good agreement with the observed abundance gradients.\\
In Figure \ref{f:gas_present} we present the comparison between the predictions of our models and the observations of the total surface gas density from \citet{Dame93} and \citet{Nakanishi03,Nakanishi06}. Concerning the data, we consider $\Sigma_{gas}=1.4($HI+H$_2)$, where the factor 1.4 accounts for the presence of He.

We note that the models adopting only radial gas flows (MW D and MW E) underestimate the gas surface density for distances $\gtrsim10$ kpc. Model MW D is also not able to reproduce the observed density in the more inner regions ($R<8$ kpc). For what concerns the models with variable SFE (MW B and MW F), they both fall inside the region covered by the observations in the inner regions. However, Model MW B overestimates the observed gas density in the more external part of the disc ($R\gtrsim14$kpc), whereas Model MW F shows a good agreement with data in the whole radial range.

\begin{figure*}
    \centering
    \includegraphics[width=1\textwidth]{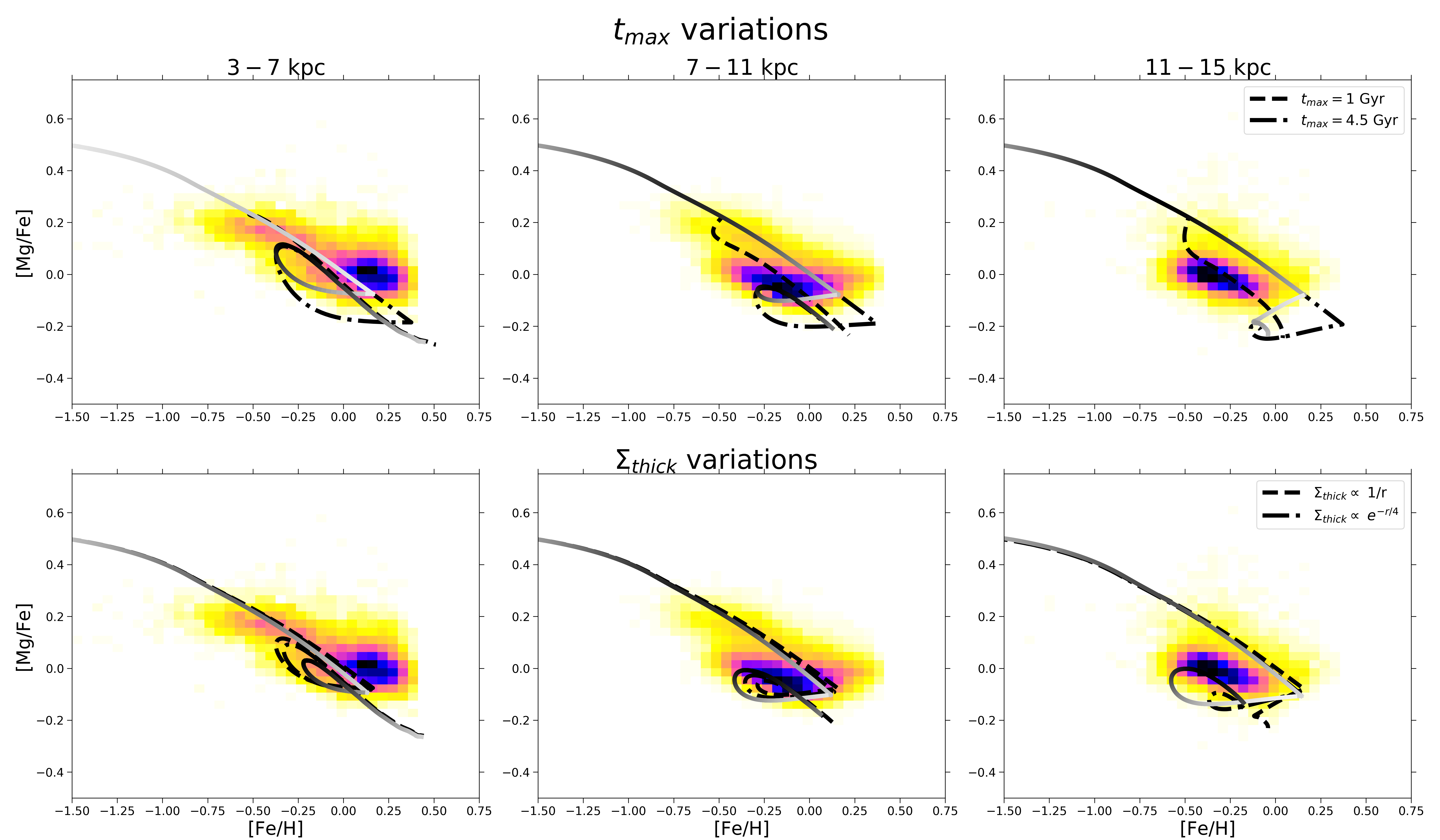}
    \caption{[$\alpha$/Fe] vs. [Fe/H] abundance ratios for different prescriptions of $t_{max}$ and $\Sigma_{thick}$. Upper panels show the effects of $t_{max}$ variations: Model MW F ($t_{max}=1\,$Gyr,  black dashed line), Model MW F1 ($t_{max}=3.25\,$Gyr, solid colour coded as in Figure \ref{f:MgFe_bestgradient}) and  Model MW F2 ($t_{max}=4.5\,$Gyr, black dash-dotted). Lower panels show the effects of $\Sigma_{thick}$ variations: Model MW F1 ($\Sigma_{thick}\propto1/R$, black dashed), Model MW F3 ($\Sigma_{thick}\propto e^{-R/4}$, black dash-dotted) and  Model MW F4 ($\Sigma_{thick}\propto e^{-R/2.3}$, solid  colour coded as in Figure \ref{f:MgFe_bestgradient}).
    Data are from \citet{Hayden15}. Left, central and right panels show the models and data in the ranges 3 < $R/$kpc <7, 7 < $R/$kpc < 11 and 11 < $R/$kpc < 15, respectively.}
    \label{f:MgFe_variations}
\end{figure*}

\subsection{The [$\alpha$/Fe] vs. [Fe/H] relation along the thin disc}
\label{ss:alpha_results}
As a second step, we then consider only Models MW F and MW E, which are able to well reproduce the abundance gradients as well as the SFR gradient, to predict the behaviour of the  [$\alpha$/Fe] vs. [Fe/H] at various Galactocentric radii, and compare them with APOGEE data (\citealt{Hayden15}). In this way, we can test if our models are able not only to reproduce the solar vicinity situation, but also the evolution of abundances at different radii.\\ 

The abundance ratio patterns are shown for Model MW F and MW E in Figure \ref{f:MgFe_bestgradient}, where colour-codes for models and data represent the stellar number density at a given model timestep and the normalised number of observed stars in certain [Mg/Fe] vs. [Fe/H] bin, respectively. Such colour-coding helps in understanding the compatibility between models and data. In fact, regions with higher data density should broadly correspond to higher stellar density predictions by the models.\\
In Figure \ref{f:MgFe_bestgradient} the predicted abundance ratios computed for three different radii (4 kpc, 8 kpc and 12 kpc) are shown and compared with data of corresponding disc regions (3-7 kpc, 7-11 kpc and 11-15 kpc).

\begin{figure}
    \centering
    \includegraphics[width=0.95\columnwidth]{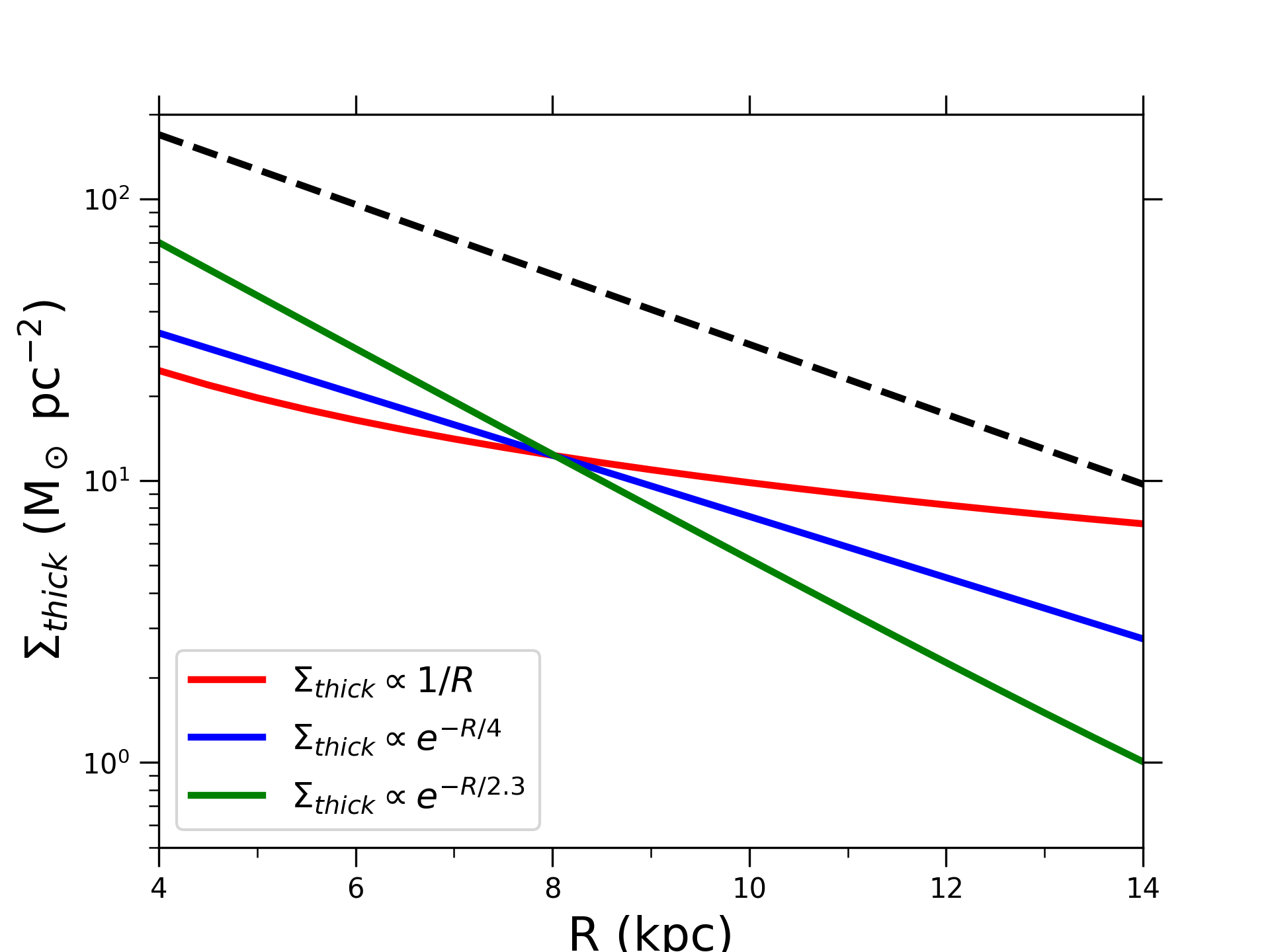}
    \caption{Assumed total surface density profiles for the thick disc adopted in this work. Solid lines are computed for $\Sigma_{thick}\propto 1/R$ (red), $\Sigma_{thick}\propto e^{-R/4}$ (blue) and $\Sigma_{thick}\propto e^{-R/2.3}$ (green). The black dashed line indicates the assumed density profile for the thin disc, $\Sigma_{thin}\propto e^{-R/3.5}$.}
    \label{f:profile_thick}
\end{figure}

We see that the data shape for [Mg/Fe] vs. [Fe/H] relation is different for different regions. In particular, we note a diagonal shape for the inner regions, where the data seems to follow a unique sequence. Moving to outer radii, the densest regions are progressively at lower [Fe/H] and the diagonal shape is no more present. In particular, in central panels we see the presence of two distinct sequences (the so called high-$\alpha$ and low-$\alpha$ sequences, e.g. \citealt{Bovy16,Spitoni19}) while for the outer regions we see only stars with low [$\alpha$/Fe].\\
For what concerns the models, they exhibit the typical patterns of the "classical" two-infall model, with the gap marking the transition between the thick and thin disc phases (see \citealt{Grisoni17,Grisoni18} for more details). This gap is not seen for Model MW F at 4 kpc due to the high SFE in the second infall phase that produce an overabundance of $\alpha$-elements (for the "time-delay model": see \citealt{Matteucci03,Matteucci12}).\\
The comparison between data and models shows that models fail in reproducing the data, especially in the outer and inner regions. In fact, in the left panels (3< r/kpc < 7) we see that the model tracks fail to pass in the densest data region. Regarding the right panels (11< r/kpc <15), model MW F in particular is able to cover the densest regions of the diagram; however it seems to predict a too low number of stars in correspondence of this abundance region. Moreover, it is clear from both the right panels that the models predict a significant amount of stars forming at high [$\alpha$/Fe] and low [Fe/H], which are not observed.

\subsubsection{The effects of $t_{max}$ and $\Sigma_{thick}$ variations}

Due to the difficulties encountered in reproducing the data at different radii, we run models varying the $t_{max}$ (time for the maximum mass accretion onto the thin disc) and  $\Sigma_{thick}$ (surface mass density profile in the thick disc) parameters. The effects of the variation of these two parameters are shown in Figure \ref{f:MgFe_variations}, where the results of Model MW F, F1, F2, F3 and F4 are plotted. We decide to avoid the plotting of MW E (E1, E2, etc.) models, since the effects of these parameter variations are very similar.\\

In the upper panels of Figure \ref{f:MgFe_variations}, we see that the agreement between model results and data in the region between 7 and 11 kpc slightly improves if we assume a longer time for maximum infall on the thin disc (i.e. $t_{max}$ from 1 to 4.5 Gyr), and this corresponds to enlarging the time during which there is a gap in the star formation between the assembly of the two discs. In particular, the model with $t_{max}=$3.25 Gyr  (MW F1) fits quite nicely the high density regions of stars at Galactocentric distances between 7 and 11 kpc. Larger values of $t_{max}$ produce instead a too large loop which does not agree with the the density distribution of stars.
On the other hand, at outer radii the models with larger $t_{max}$ completely fail to pass for the region covered by the largest number of stars. At inner radii instead, a longer $t_{max}$ creates a loop which is not consistent with the observed diagonal sequence of denser data regions.\\
Our modification in the second infall delay time is justified by results found in previous works. In fact, a long value for $t_{max}$ was suggested by \citet{Spitoni19,Spitoni20} who well reproduced chemical abundances and stellar ages in the solar neighbourhood from the APOKASC sample (\citealt{Silva18}) by imposing a large delay ($\sim$ 4 Gyr) in the second infall episode. 

\begin{figure*}
    \centering
    \includegraphics[width=1.\textwidth]{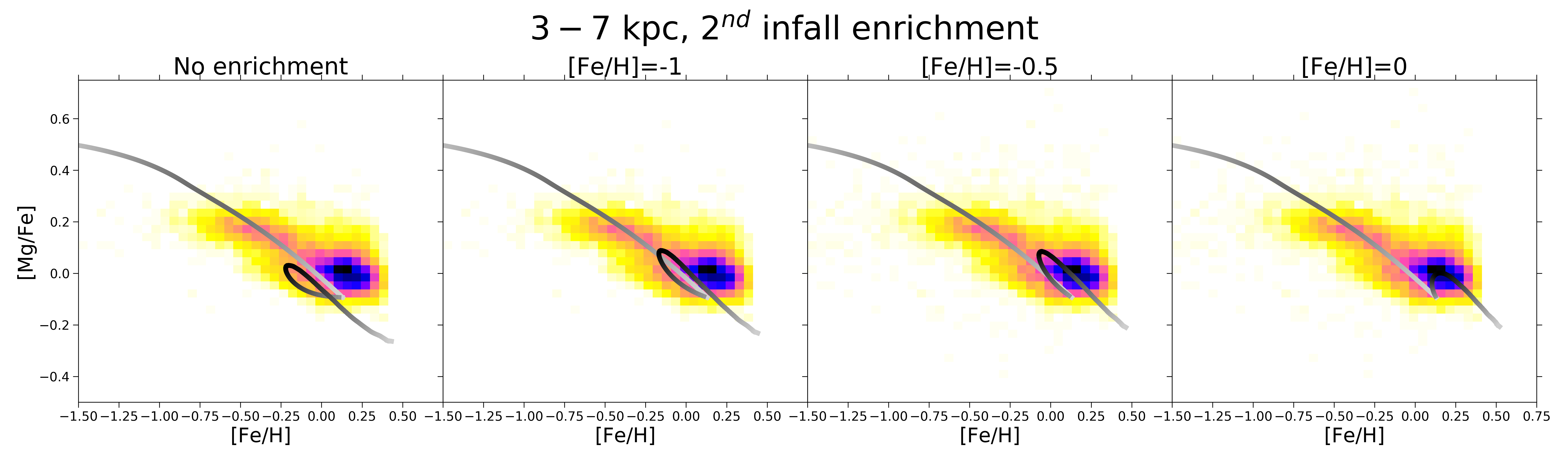}
    \caption{Effects on the chemical evolution at 4 kpc in the [$\alpha$/Fe] vs. [Fe/H] abundance ratios in varying the enrichment of the second infall episode.  Panels show the models with pre-enrichment (from left to right): primordial (Model MW F4), [Fe/H]=-1 (Model MW F5), [Fe/H]=-0.5 (Model MW F6), [Fe/H]=0 (Model MW F7). Data are from \citet{Hayden15}.}
    \label{f:FeH_variation}
\end{figure*}

In Figure \ref{f:MgFe_variations}, we can see also the effect of assuming a $\Sigma_{thick}$ variable with Galactocentric distance, but in a different way than $\propto 1/R$ (see Model MW F3 and MW F4 in Table \ref{t:models}).  Recent works proposed a very centrally peaked density for the thick disc (\citealt{Bensby11,Bovy12,Haywood16}), with scale length of the order of $2$ kpc (e.g. \citealt{Pouliasis17}). The variation in the thick disc profile reflects in the ratio between thin and thick disc densities ($\Sigma_{thin}/\Sigma_{thick}$), which can have a great influence on the chemical evolution patterns at different radii.\\
This can be clearly seen in the lower panels of Figure \ref{f:MgFe_variations}, where we test three different surface density profiles for the thick disc: in particular, we look at the profiles $\Sigma_{thick}\propto1/R$ (\citealt{Colavitti09,Romano10}, model F1), $\Sigma_{thick}\propto e^{-R/4}$ (\citealt{Cautun19}, model F3) and $\Sigma_{thick}\propto e^{-R/2.3}$ (\citealt{Pouliasis17}, model F4). We remind that we fix $\Sigma_{thin}(8$kpc$)/\Sigma_{thick}(8$kpc$)\sim 4$ for all the models, according to \citet{Spitoni20}.  The different profiles are plotted in Figure \ref{f:profile_thick}, together with the density profile of the thin disc (dashed line in the Figure).

In the lower panels of Figure \ref{f:MgFe_variations}, we observe that a higher $\Sigma_{thin}/\Sigma_{thick}$ ratio produces more evident loops. This allows the model with $\Sigma_{thick}\propto e^{-R/2.3}$  to reproduce the data density distribution at $R>11$ kpc. The obtained result suggests a ratio $\Sigma_{thin}/\Sigma_{thick}\sim 8$ in the outer disc regions, hence a ratio increasing with radius, in agreement with what noted in \citet{Anders14}. This is exactly the opposite of what is found with "classical" prescriptions, where the ratio in the outer regions is halved rather than doubled relative to the solar neighbourhood (see Figure \ref{f:profile_thick}). In the inner regions ($R<7$kpc), the most centrally peaked thick disc profile contributes to alleviate the problem of the prominent loop. This latter feature still prevents to cover the region with the highest density of stellar data.\\

However, the Model MW F4 remains the best among the models shown in Figure \ref{f:MgFe_variations}, due to the better agreement found for the central (7-11 kpc) and outer (11-15 kpc) regions of the MW. Moreover, this model maintains a good agreement with present-day abundance and SFR gradients: the variations resulted from the different $t_{max}$ and $\Sigma_{thick}(R)$ are in fact limited ($\lesssim 0.015$ dex/kpc in abundance gradients slopes).

\subsubsection{The effects of enriched gas infall}

Because of the problems found for the [$\alpha$/Fe] ratios, we want to further investigate the evolution of the inner regions. In particular, we test whether a metal enriched second infall can explain the observed behaviour. This is done by adopting a second infall with abundances  that we obtain from the model of the thick disc corresponding to some specific [Fe/H] values.
The adoption of the abundance patterns of the thick disc does not mean that the gas is enriched only by the gas lost from the thick disc, but rather that this enriched infall is due partly to the gas lost from the formation of the thick disc, Galactic halo or the Galactic bar which then gets mixed with a larger amount of infalling primordial gas. We should note that this was already proposed by \citet{Gilmore86}. However, the present paper can address data sets which were not available at that time, thus providing more stringent constraints.

The results of such analysis are shown in Figure \ref{f:FeH_variation}. In particular, we test three levels of pre-enrichment for the second infall: [Fe/H]=-1,-0.5 and 0 dex (Models MW F5, F6 and F7). Lower metallicities for the infall imply too little variations relative to a primordial infall; moreover, such metallicities are reproduced by halo models only at $t\lesssim$1 Gyr (e.g. \citealt{Spitoni16}). A supersolar infall, instead, would produce too large abundance values at present time, which are not consistent with present-day gradient observations.

\begin{figure*}
    \centering
    \includegraphics[width=1.\textwidth]{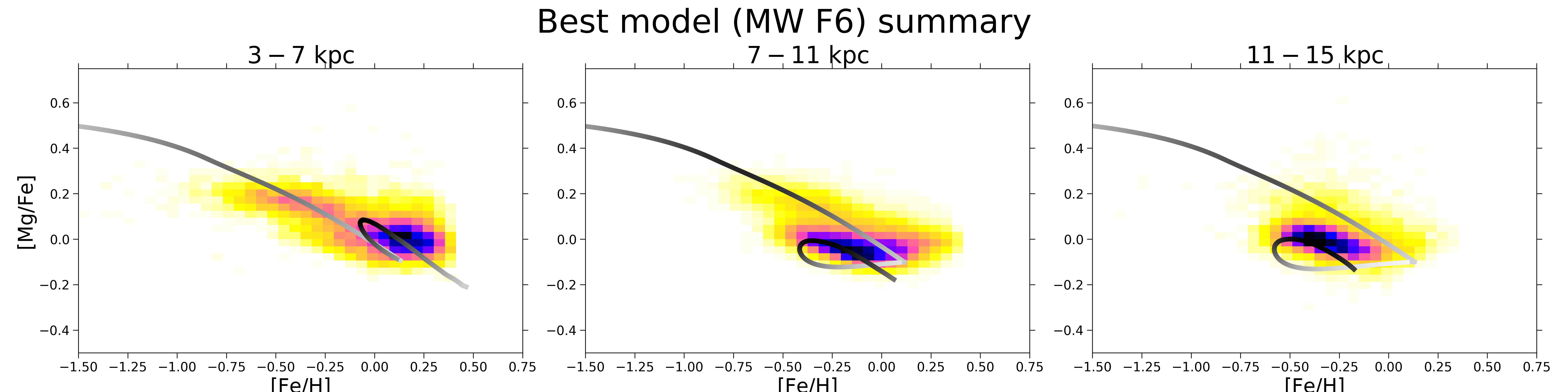}\\
    \includegraphics[width=1.\textwidth]{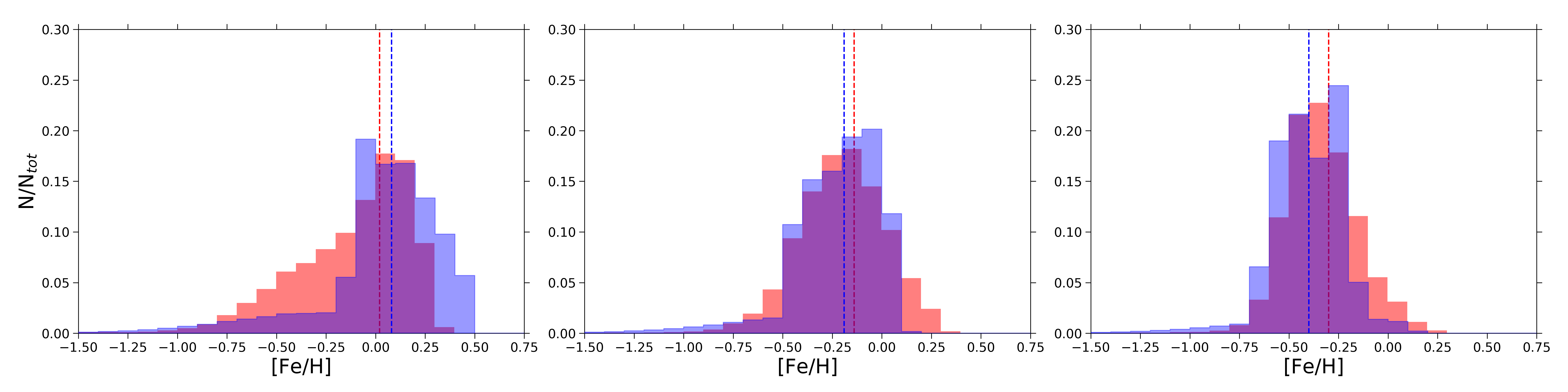}\\
    \includegraphics[width=1.\textwidth]{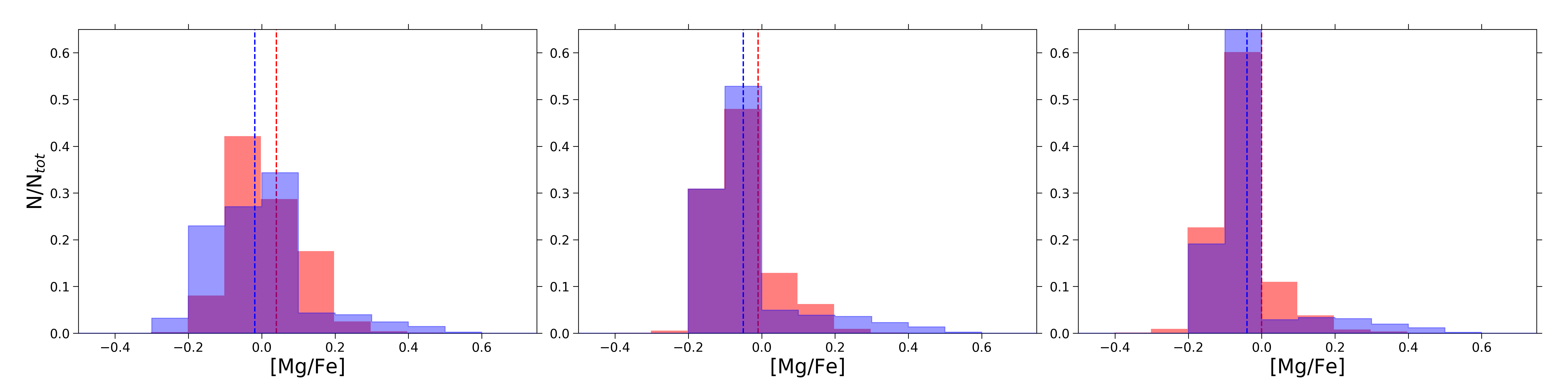}
    \caption{Summary for the resulting best model MW F6. Data are from \citet{Hayden15}. Left, central and right panels show the models and data in the ranges 3 < $R/$Kpc <7, 7 < $R/$kpc < 11 and 11 < $R/$Kpc < 15, respectively. Upper panels: [$\alpha$/Fe] vs. [Fe/H] abundance ratios. Middle panels: [Fe/H] MDFs. Lower panels: [Mg/Fe] MDFs. Blue and red dashed lines are the median [X/Y] ratios for models and data, respectively.}
    \label{f:summary_bestmodel}
\end{figure*}

From Figure \ref{f:FeH_variation}, we see that a significant gas enrichment is necessary to explain APOGEE data of the thin disc in the inner regions. Going more into detail, we obtain a good agreement for Model MW F6, with [Fe/H]=-0.5 dex enriched infall. In fact, the high [Fe/H] enrichment prevents a too low [Fe/H] at the beginning of the thin disc accretion, while the high [$\alpha$/Fe] in the infall gas boosts the model track to higher [Mg/Fe] values in the thin disc.\\
At a first sight, the Model MW F7 ([Fe/H]=0 dex) seems to have an even higher data-model agreement, with densest data region covered by the model. However, the high level of enrichment in the second infall totally prevents to explain the large number of data at moderate metallicity (-0.2 dex<[Fe/H]<0.1 dex, see the inner region data MDF in Figure \ref{f:summary_bestmodel}).

\subsubsection{Discussion}

In Figure \ref{f:summary_bestmodel} we summarise the results for the model that we consider the best (MW F6) in reproducing the data. This model assumes a variable SFE, radial gas flows with constant speed, a time for the maximum infall onto the disc of 3.25 Gyr, an enriched infall ($[Fe/H]=-0.5$ dex) and a distribution of the total surface mass density of the thick disc $\Sigma_{thick}\propto e^{-R/2.3}$. In particular, we show the [Mg/Fe] vs. [Fe/H] plots (upper panels), the [Fe/H] MDFs (middle panels) and the [Mg/Fe] MDFs (lower panels); we also plot the median $<$[Fe/H]$>$ and $<$[Mg/Fe]$>$ resulting from data (red dashed line) and models (blue dashed line). The median abundances, together with their uncertainties, are presented in Table \ref{t:DF_bestmodel}. We point out that model MDFs and median abundances in each disc region (3-7 kpc, 7-11 kpc, 11-15 kpc) are calculated considering the results of all the rings included in each region (4 kpc and 6 kpc in the range 3-7 kpc; 8 kpc and 10 kpc in the range 7-11 kpc; 12 kpc and 14 kpc in the range 11-15 kpc).

We find model MW F6 as the best in reproducing the data features at different radii, as shown in Figure \ref{f:summary_bestmodel}. In fact, as can be seen from the upper panels, this model predicts [Mg/Fe] ratios in good agreement with data, with the peaks in the data density nicely reproduced by the predicted stellar number densities. This is also visible from the [Fe/H] MDFs in the central row of panels of Figure \ref{f:summary_bestmodel}. 

The model [Fe/H] MDFs at median and outer radii in general reproduce very well the observational data, although the number of metal rich ([Fe/H]$\gtrsim$0 dex) stars is slightly underestimated at radii > 11 kpc. This underestimation may have been caused by radial migration of stars (e.g. \citealt{Schonrich09,Minchev18}). In fact, \citet{Vincenzo20} showed that a non negligible fraction of outer radii stars may have been born in the Galaxy inner regions (see Figure 8 \citealt{Vincenzo20}). However, \citet{Vincenzo20} also found that in general less than 10\% of stars have been migrated for more than 2 kpc.\\
In the inner radii the model predicts a too low number of stars with metallicity [Fe/H]$\leq$-0.2 dex. This latter feature is more or less common to all the models with delayed second infall (i.e. greater $t_{max}$). This low number of low metallicity stars can be possibly solved by adopting an even denser thick disc profile for the inner regions (model MW F6 already adopts the most centrally peaked profile tested). Also stronger radial gas flows in the inner regions can help in alleviate the observed discrepancy, although to a lesser extent. 

\begin{table}
    \centering
    \caption{Median abundance ratios obtained from APOGEE data of \citet{Hayden15} and the best model MW F6. In the first column, we write the elements and the source considered (model or data) . In the second, third and fourth column we list the obtained median abundance ratios (with $\pm \sigma
    $ errors) in the radius ranges 3-7 kpc, 7-11 kpc and 11-15 kpc, respectively.} 
    \begin{tabular}{c | c  c  c }
    \hline
    \hline
     & 3-7 kpc  & 7-11 kpc  &   11-15 kpc \\
    \hline
    $<$[Fe/H]$>$ data & 0.02$^{+0.21}_{-0.39}$  & -0.14$^{+0.22}_{-0.22}$ & -0.30$^{+0.20}_{-0.16}$ \\[0.2cm]
    $<$[Fe/H]$>$ model & 0.08$^{+0.22}_{-0.21}$ & -0.19$^{+0.17}_{-0.21}$ & -0.40$^{+0.18}_{-0.17}$ \\[0.1cm]
    \hline
    $<$[Mg/Fe]$>$ data & 0.04$^{+0.13}_{-0.07}$ & -0.01$^{+0.10}_{-0.06}$ & 0.00$^{+0.07}_{-0.05}$ \\[0.2cm]
    $<$[Mg/Fe]$>$ model & -0.01$^{+0.10}_{-0.13}$ & -0.05$^{+0.05}_{-0.08}$ & -0.04$^{+0.04}_{-0.06}$ \\[0.1cm]
    \hline
    \hline
\end{tabular}
\label{t:DF_bestmodel}
\end{table}

Concerning [Mg/Fe] MDFs, we have a good agreement from solar to outer radii (7-15 kpc), where the sharp peaks in the data MDFs is exactly reproduced.\\
The broader MDF observed in the inner radii is also predicted by the model. However, the model MDF seems to show an offset of $\sim$0.1 dex relative to the observed one. This offset (which is roughly compatible with APOGEE uncertainties) can be in part attributed to the model underestimation of low [Fe/H] stars. On the other hand, some recent works claimed some systematics in APOGEE abundance determinations (J\"onsson et al. 2020, submitted). In particular, such problems seem to affect $\alpha$-elements, where too large [$\alpha$/Fe] are observed at high [Fe/H] (\citealt{Matteucci20}, see also \citealt{SantosPeral20}).

In any case, the general agreement between model results and data is confirmed by Table \ref{t:DF_bestmodel}. The median abundances obtained from the models are in fact well within the 1$\sigma$ range obtained from the data.\\

\begin{figure}
    \centering
    \includegraphics[width=1\columnwidth]{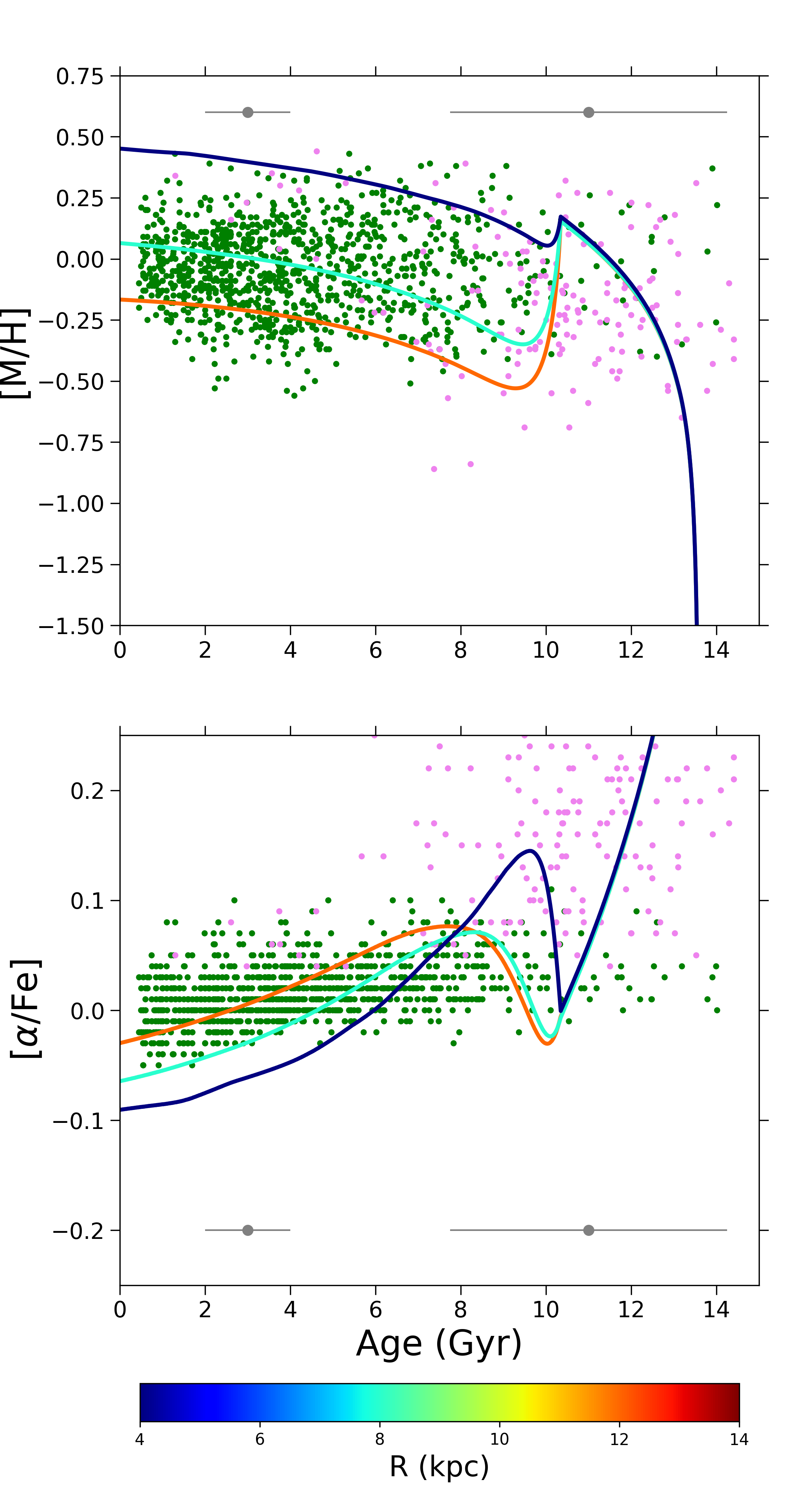}
    \caption{Time evolution of [M/H] (upper panel) and  [$\alpha$/Fe] (lower panel) ratios for the stellar sample presented by \citet{Silva18}, compared with the predictions of the best model MW F6. Pink points depict the high-$\alpha$ population defined as in \citet{Silva18}, whereas green points represent the low-$\alpha$ one. As in \citet{Spitoni19,Spitoni20}, we have not taken into account Y$\alpha$R stars. The grey symbols in the two panels indicate median age uncertainties for 3 and 11 Gyr old stars of the sample. The bottom colorbar indicates the radius at which each line is computed.}
    \label{f:AgeMet_data}
\end{figure}

We also test the dependence on radius of the age-metallicity relation for the best model MW F6.\\
This is shown in Figure \ref{f:AgeMet_data}, where we see the time evolution of [M/H] (upper panel) and [$\alpha$/Fe] (lower panel) at different radii. To probe the robustness of our best model, we compare the predictions to abundances and ages from the updated APOKASC sample by \citet{Silva18}. In that paper, the authors chemically separated the high-$\alpha$ and low-$\alpha$ disc populations. They used a sample of red giant stars spanning out to $\sim$2 kpc in the solar annulus and provided precise stellar ages by means of asteroseismology. In order to be consistent with available abundances associated to asteroseismic data, we compute the metallicity adopting the expression introduced by \citet{Salaris93}:
\begin{equation}
[M/H]=[Fe/H]+\log(0.638\times10^{[\alpha/Fe]}+0.362)
\end{equation}
where for [$\alpha$/Fe] we mean [Si+Mg/Fe].

In Figure \ref{f:AgeMet_data} upper panel, we note that the chemical evolution tracks show progressively decreasing [M/H] with radius during the second infall phase. This behaviour is caused by the increasing SFE with decreasing radius and radial gas flows. At the same time, we see that for the most internal radii (4 kpc) the drop in metallicity between the end of the first infall and the beginning of the second one vanishes due to the enriched infall.\\
Our best model at 8 kpc generally reproduces the trend of both high and low-$\alpha$ data. The other two lines of Figure \ref{f:AgeMet_data} upper panel can be regarded as genuine predictions by the best model. In fact, the guiding radii of the low-$\alpha$ sample are comprised between $\sim$6 and $\sim$9 kpc (see Figure 13 of \citealt{Silva18}). Nonetheless, the suggestion of an enriched infall for the most internal radii ($R<8$kpc) could explain the most metal rich stars.

Concerning the lower panel of Figure \ref{f:AgeMet_data}, we note in this case that our model well fits the age of the high-$\alpha$ sequence stars, as defined in \citet{Silva18} (pink points). The model is also able to explain the relatively young ($\sim6-8$ Gyr), high-$\alpha$ stars as a consequence of the enriched second-infall assumed for the most internal radii. In fact, the high-$\alpha$ sample exhibits lower average radii than the low-$\alpha$ one ($\sim$6 kpc instead of $\sim$8 kpc).\\ 
At the same time, the model reproduces the bulk of low-$\alpha$ data, although the [$\alpha$/Fe] ratio in young stars seems be to be understimated by the model. However, we already pointed out that APOGEE (and so also APOKASC) abundance may overstimate the [$\alpha$/Fe] ratios at high metallicities (\citealt{Matteucci20}).

\subsection{Enceladus}

Since we tested that enriched infall allows us to better fit APOGEE data in the inner Galactic thin disc regions (3-7 kpc), we investigate the possibility that the infall episodes are related to the merging event occurred in the inner halo and called Gaia-Enceladus (\citealt{Helmi18}). In particular, we test if some Enceladus gas can contribute to the first or the second infall episode.\\
To see whether Enceladus plays a significant contribution to MW disc evolution, we run a chemical evolution model for Enceladus itself (see Section \ref{ss:Enceladus_model}) adopting the parameters of the best model of \citet{Vincenzo19} (see Table \ref{t:Enceladus_param}).\\

\begin{figure}
    \centering
    \includegraphics[width=0.95\columnwidth]{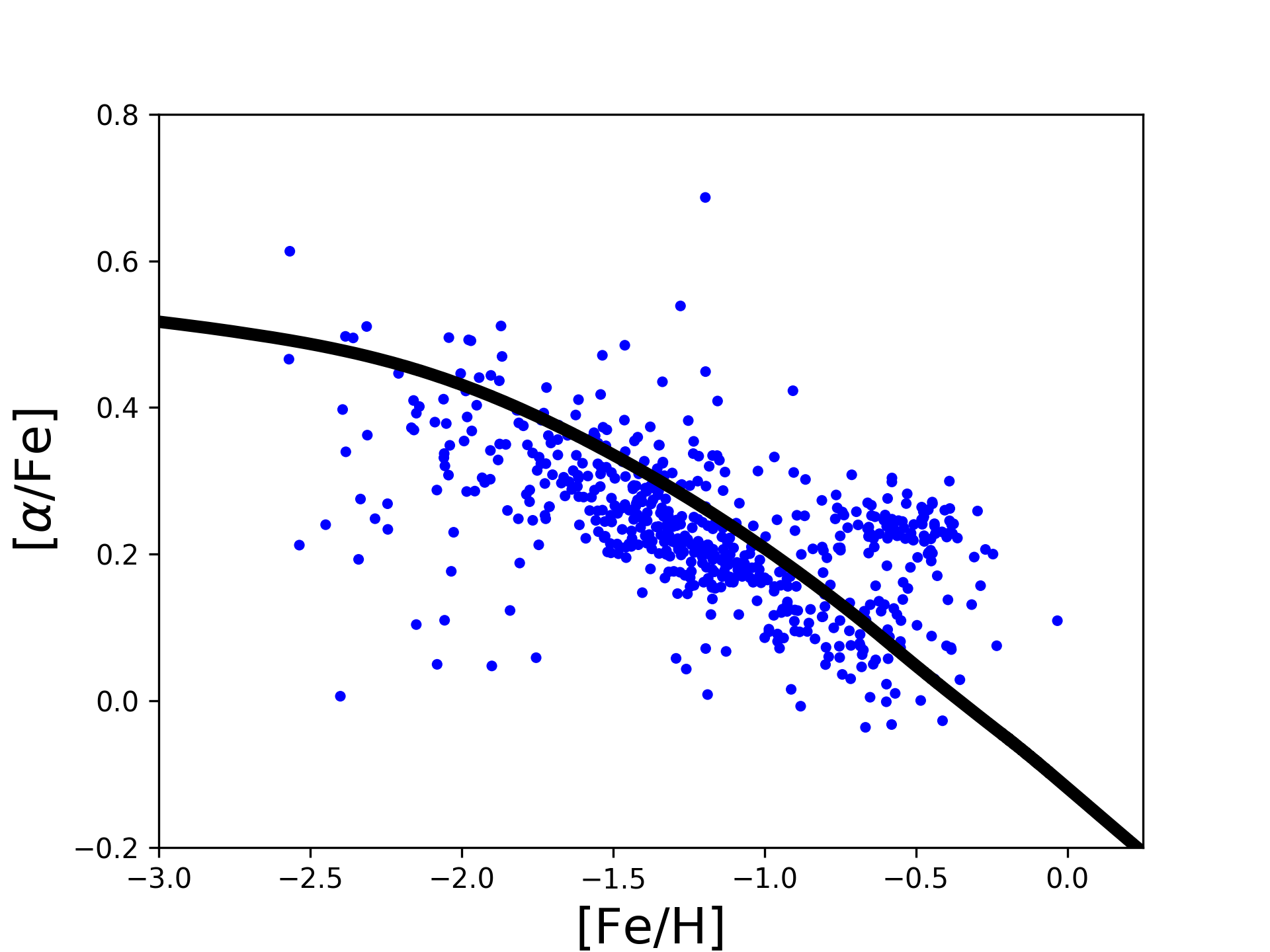}
    \caption{[$\alpha$/Fe] vs. [Fe/H] abundance ratios for Enceladus chemical evolution model (black solid line) compared with data from \citet{Helmi18} (blue dots).}
    \label{f:Enceladus}
\end{figure}

We test a possible contribution to the second infall episode from Enceladus by assuming its measured gas abundances at the start of the second infall  assumed to occur at $3.25$ Gyr, as in model MW F4. Since our MW models start $13.6$ Gyr ago, the above assumptions on the infall gas are reasonable: in fact, \citet{Helmi18} placed the merger approximately $10$ Gyr ago, while in \citet{Chaplin20} they found that the 68\% confidence upper age limit was of 11.6 Gyr. \\
As can be seen from Figure \ref{f:Enceladus}, the predicted [$\alpha$/Fe] ratios for Enceladus which agree very well with the stellar data from \citealt{Helmi18}, show lower values than found in the MW, as suggested by the "time-delay model" (see \citealt{Matteucci03,Matteucci12}). In this formalism, if the SFR of Enceladus is lower than in the solar vicinty (and in MW inner radii), we expect a steeper decrease of the [$\alpha$/Fe] ratios as functions of [Fe/H]. The evolution of Enceladus is in fact assumed to be similar to that of a dwarf spheroidal galaxy.

\begin{figure}
    \centering
    \includegraphics[width=1\columnwidth]{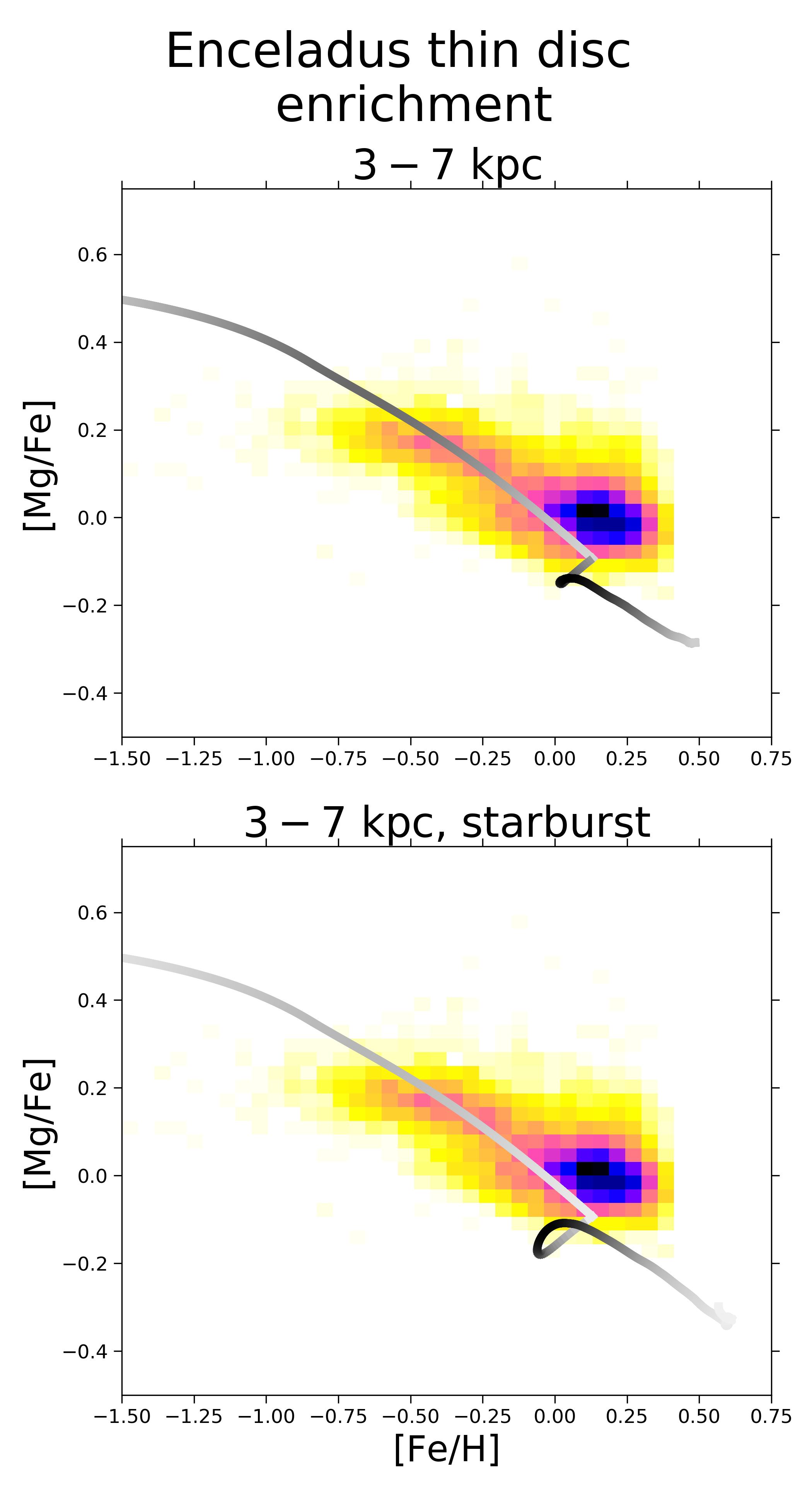}
    \caption{Effects on the chemical evolution at 4 kpc in the [$\alpha$/Fe] vs. [Fe/H] abundance ratios in adopting a second infall enriched by Enceladus gas at 15\%. All panels refer to MW F4 based models. Upper panel: results for standard infall timescales. Lower panel: results for starburst ($\tau=1\,Gyr$). Data are from \citet{Hayden15}.}
    \label{f:enrich_Enceladus_thin}
\end{figure}

The lower [$\alpha$/Fe] ratios of Enceladus gas prevent the agreement between MW models with second infall enriched by Enceladus and data, even adopting a partial contribution to the enrichment of the infalling gas. 
In fact, the low Enceladus [Mg/Fe] in the infalling gas (which does not vary significantly changing the contribution of Enceladus to the infall) causes a significant lowering in the [Mg/Fe].
This is shown in Figure \ref{f:enrich_Enceladus_thin} upper panel, where 15\% of the second infall is constituted by Enceladus enriched gas. We choose this contribution from Enceladus in order to have a consistent picture in the mass budget. In fact, our Enceladus model predicts a $M_{gas}\sim 2\cdot 10^9$ M$_\odot$ at $t=3.25$ Gyr, which is $\simeq 15$ \% the mass accreted during the second infall episode by inner regions (3-7 kpc). We highlight that our result is robust against the different mass estimates for Gaia-Enceladus (\citealt{Helmi18,Vincenzo19}: $\sim 10^9$ M$_\odot$, \citealt{Mackereth20}: $\sim 3 \cdot 10^8$ M$_\odot$ ). Even limiting the enriched accretion within 1 Gyr from the start of the second infall (coherently with \citealt{Mackereth20} mass estimate), we obtain results inconsistent with data. Inconsistent results are also obtained by adopting a higher Enceladus contribution to the infall (e.g. 25\%, 50\%).

Recent papers (e.g. \citealt{Grand20}) also suggest that the Enceladus merger contribute 10-50\% of gas to a centrally concentrated starburst. 
We simulate the Enceladus induced starburst by reducing significantly the second infall timescale in the inner regions of the galaxy ($< 7$ kpc). In particular, we adopt $\tau_2=1$ Gyr and an infall pattern where Enceladus gas contributes for 15\%.
However, as it can be seen in Figure \ref{f:enrich_Enceladus_thin} lower panel,  the very short infall timescale does not solve the problem of too low [Mg/Fe] ratios predicted by the model in the upper panel.\\
Similar conclusions are reached also varying the input parameters for the Enceladus enriched infall. In particular, we test what happens with a second infall starting at 2 Gyr, i.e. coincident with the age upper limit suggested by \citet{Chaplin20} for the start of Enceladus accretion to the Galaxy. We also try to change the infall patterns from Enceladus gas (e.g. [Fe/H]=-1 dex). However, all these models are not able to explain the observed data trend.\\

\begin{figure}
    \centering
    \includegraphics[width=1\columnwidth]{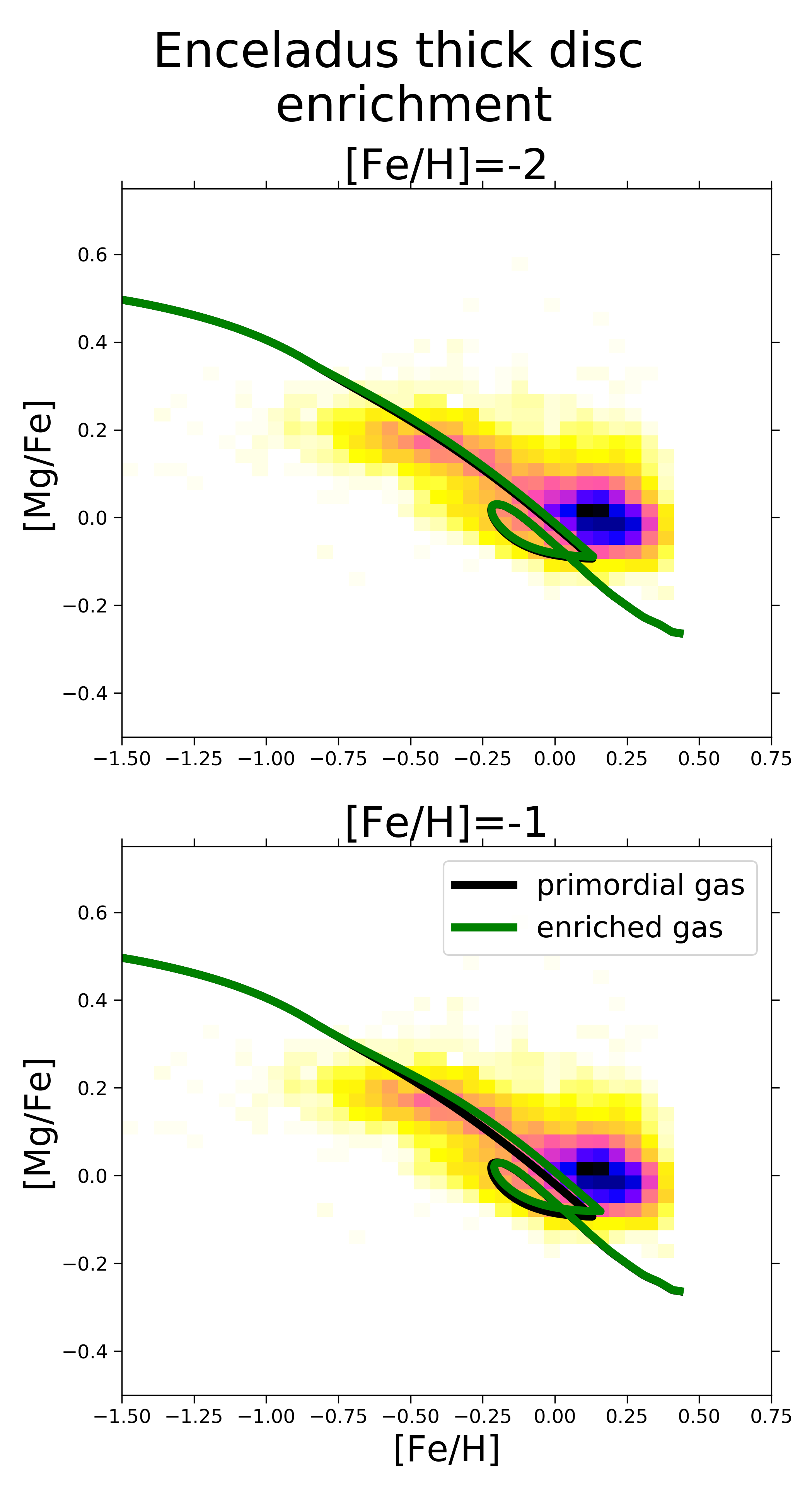}
    \caption{Effects on the chemical evolution at 4 kpc in the [$\alpha$/Fe] vs. [Fe/H] abundance ratios in adopting a first infall fully enriched by Enceladus gas from $t>0.5$ Gyr . All panels refer to MW F4 based models. Upper panel: results for Enceladus gas at [Fe/H]=-2 dex. Lower panel: results for Enceladus gas at [Fe/H]=-1 dex. Data are from \citet{Hayden15}.}
    \label{f:enrich_Enceladus_thick}
\end{figure}

We also explore the possibility that Enceladus gas contributes to the first infall giving rise to the thick disc.
To this scope, we try different levels of enrichment ([Fe/H]=-2,-1,-0.5 dex, adopting the abundance patterns from Enceladus model in Figure \ref{f:Enceladus}), as well as different times for the start of the infall enrichment (0, 0.5, 1, 2 Gyr).

In Figure \ref{f:enrich_Enceladus_thick}, we show the results for a thick disc infall enrichment starting at 0.5 Gyr. In the upper panel, we see that a low metal enriched gas ([Fe/H]=-2 dex) does not bring any evident change in the predicted abundance track. Concerning the lower panel, the moderate enrichment (1/10 solar in iron) has little effect in enhancing the [$\alpha$/Fe] ratio but is still not able to explain the abundance trend.

\section{Conclusions}
\label{s:conclusions}
In this paper we have computed the chemical evolution of the thick and thin discs of the Milky Way by adopting a two-infall model paradigm. In other words, we have assumed that the thick and thin discs form by means of two infall episodes separated by a period in which star formation decreases and then increases again thanks to the second infall episode.\\

We run several models by varying physical parameters such as the presence or absence of radial gas flows and constant or variable efficiency of star formation along the thin disc and we have compared the results of our models with abundance and SFR gradients along the thin disc, selecting the best models to reproduce the gradients.\\
Then we varied the duration of the interval between the formation of the thick and thin discs, expressed as the time of maximum gas infall onto the disc, the total present time surface gas density of the thick disc, and finally the thin disc infalling gas chemical composition in the inner radii. For all the models, we did not change the IMF, the inside-out scheme for the formation of the thin disc, the timescale of the infall and the efficiency of star formation for the thick disc. We showed the predictions of these models concerning the [Mg/Fe] vs. [Fe/H] relations along the disc compared with APOGEE data.  We also computed the metallicity distribution functions at different Galactocentric radii, selecting the best model in reproducing the majority of the observational constraints.

Moreover, we computed the evolution of a dwarf galaxy resembling Gaia-Enceladus and adopted the predicted abundance pattern, which agrees with observations, as the chemical composition of the infalling gas either for the thick or thin disc.\\

Our main conclusions can be summarized as follows:
\begin{enumerate}
    \item It is not possible to reproduce the observed abundance gradients along the thin disc only assuming an inside-out formation as the main responsible mechanism for gradient formation. We need to assume either radial gas flows or efficiency of star formation varying with the Galactocentric distance, in agreement with \citet{Grisoni18}. However, too strong radial gas flows produce too steep abundance and gas density gradients and models adopting only variable SF efficiency and no radial gas flows do not reproduce the gradient of the SFR and gas density along the thin disc. Our best models for reproducing abundance and SFR gradients are Model MW E, assuming radial flows with variable speed progressively decreasing with radius and time, and Model MW F, with constant moderate radial gas flows and variable SF efficiency.  In addition, the latter model well matches the observed gas density gradient. 
    
    \item Model MW E and MW F have a maximum timescale for infall onto the disc $t_{max}=1$ Gyr, as in the "classical" two-infall model (\citealt{Chiappini97,Grisoni17}), and this choice does not produce very good agreement with the observed [Mg/Fe] vs. [Fe/H] relations especially in the inner (3-7 kpc) and outer (11-15 kpc) Galactic regions. Better agreement is reached if $t_{max}=3.25$ Gyr, in accord with previous suggestions by \citet{Spitoni19,Spitoni20}. Results for the outer regions suggest also the adoption of a small scale length ($R_d\sim2$kpc) exponential profile for the thick disc, in agreement with recent findings (e.g. \citealt{Anders14,Haywood16,Ciuca20}).
    
    \item Even better agreement with data is reached if the infall onto the inner thin disc is enriched at the level of [Fe/H]=-0.5 dex. This can be the effect of mixing between the gas leftover from the formation of the thick disc, Galactic halo or Galactic bar and the more abundant primordial extragalactic infalling gas. Finally, after analysing the metallicity distribution functions at different radii, we conclude that the best model is MW F6, with moderate gas flows, variable SF efficiency, $t_{max}=3.25$Gyr and centrally peaked exponential distribution for the total surface density of the thick disc, plus enriched infall along the inner thin disc.\\
    However, a word of caution is necessary when speaking of the APOGEE data relative to [Mg/Fe] at high [Fe/H] values, since the [Mg/Fe] ratio  could have been overestimated, as suggested by J\"onsson et al. (2020, submitted).
    
    \item The chemical evolution of Enceladus is computed by adopting the same parameters, shown in Table \ref{t:Enceladus_param}, as in \citet{Vincenzo19} and we do reproduce the observed abundances by \citet{Helmi18}. Assuming gas infall enriched as the abundance pattern derived for Enceladus at 3.25 Gyr (starting time for the thin disc infall) fails in reproducing the [Mg/Fe] vs. [Fe/H] relation for the inner Galactic regions. Enceladus enriched infall during the thick disc formation instead does not alter the chemical evolution pattern. Therefore, we can conclude that the gas lost by Enceladus or a part of it could have concurred to the formation of the thick disc but not to that of the thin disc.
\end{enumerate}

\section*{Acknowledgements}

M.P. and F.M. acknowledges M. Schulteis for useful suggestions on the data.\\
E.S. acknowledges support from the Independent Research Fund Denmark (Research grant 7027-00096B). Funding for the Stellar Astrophysics Centre is provided by The Danish National Research Foundation (Grant agreement no.: DNRF106).\\
F.V. acknowledges the support of a Fellowship from the Center for Cosmology and AstroParticle Physics at The Ohio State University. 
The authors thank the anonymous referee for careful reading of the manuscript and useful suggestions.

\section*{Data availability}
The data underlying this article will be shared on reasonable request to the corresponding author.




\bibliographystyle{mnras}
\bibliography{radial_gradients}








\bsp	
\label{lastpage}
\end{document}